\documentclass{emulateapj}
\usepackage{apjfonts}
\usepackage{epsf}
\usepackage{psfig}
\begin{document}

\slugcomment{ApJ, accepted}
\shortauthors{J. M. Miller et al.}
\shorttitle{Disks and CCDs}

\title{On Relativistic Disk Spectroscopy in Compact Objects with X-ray CCD Cameras}

\author{J.~M.~Miller\altaffilmark{1},
        A.~D'A\`i\altaffilmark{2},
        M.~W.~Bautz\altaffilmark{3},
        S.~Bhattacharyya\altaffilmark{4},
        D.~N.~Burrows\altaffilmark{5},
        E.~M.~Cackett\altaffilmark{1},
        A.~C.~Fabian\altaffilmark{6},
        M.~J.~Freyberg\altaffilmark{7},
        F.~Haberl\altaffilmark{7},
        J.~Kennea\altaffilmark{5},
        M.~A.~Nowak\altaffilmark{3},
        R.~C.~Reis\altaffilmark{6},
        T.~E.~Strohmayer\altaffilmark{8},
        M.~Tsujimoto\altaffilmark{9}}

\altaffiltext{1}{Department of Astronomy, University of Michigan, 500
Church Street, Ann Arbor, MI 48109, jonmm@umich.edu}
\altaffiltext{2}{Dipartimento di Scienze Fisiche ed Astronomiche,
Universita di Palermo, Italy}
\altaffiltext{3}{Kavli Institute for Astrophysics and Space Research,
MIT, 77 Massachusetts Avenue, Cambridge, MA 02139}
\altaffiltext{4}{Department of Astronomy and Astrophysics, Tata
Institute of Fundamental Research, Mumbai 400005, India}
\altaffiltext{5}{Department of Astronomy \& Astrophysics, Pennsylvania
  State University, 525 Davey Lab, College Park, PA, 16802}
\altaffiltext{6}{Institute of Astronomy, University of Cambridge,
Madingley Road, Cambridge CB3 OHA, United Kingdom}
\altaffiltext{7}{Max-Planck-Institut f{\"u}r extraterrestrische Physik,
  Giessenbachstra{\ss}e, 85748 Garching, Germany}
\altaffiltext{8}{Astrophysics Science Division, NASA Goddard Space
Flight Center, Greenbelt, MD, 20771}
\altaffiltext{9}{Japan Aerospace Exploration Agency, Institute of
  Space and Astronomical Sciences, 3-1-1 Yoshino-dai, Sagamihara,
  Kanagawa 229-8510, Japan}


\begin{abstract}
X-ray charge-coupled devices (CCDs) are the workhorse detectors of
modern X-ray astronomy.  Typically covering the 0.3--10.0~keV energy
range, CCDs are able to detect photoelectric absorption edges and K
shell lines from most abundant metals.  New CCDs also offer
resolutions of 30--50 (E/$\Delta$E), which is sufficient to detect
lines in hot plasmas and to resolve many lines shaped by dynamical
processes in accretion flows.  The spectral capabilities of X-ray CCDs
have been particularly important in detecting relativistic emission
lines from the inner disks around accreting neutron stars and black
holes.  One drawback of X-ray CCDs is that spectra can be distorted by
photon ``pile-up'', wherein two or more photons may be registered as a
single event during one frame time.  We have conducted a large number
of simulations using a statistical model of photon pile-up to assess
its impacts on relativistic disk line and continuum
spectra from stellar-mass black holes and neutron stars.  The
simulations cover the range of current X-ray CCD spectrometers and
operational modes typically used to observe neutron stars and black
holes in X-ray binaries.  Our results suggest that severe photon
pile-up acts to falsely narrow emission lines, leading to
falsely large disk radii and falsely low spin values.  In contrast,
our simulations suggest that disk continua affected by severe pile-up
are measured to have falsely low flux values, leading to falsely small
radii and falsely high spin values.  The results of these simulations
and existing data appear to suggest that relativistic disk
spectroscopy is generally robust against pile-up when this effect is
modest.  \end{abstract}

\section{Introduction}

X-ray observations of accreting neutron stars and black holes
necessarily probe regions close to the compact object.  The long--held
promise of these observations is that aspects of the compact object
itself, and the innermost accretion flow, may be revealed.  Notable
early attempts to realize this promise took several forms.  For
instance, many attempts were made to use the blackbody spectra of
Type-I X-ray bursts in accreting neutron star systems to infer the
stellar radius; however, derived radii were implausibly small, likely
due to scattering effects (see London, Taam, \& Howard 1986).  In the
case of black holes, careful efforts to measure the inner radius of
the accretion disk (see, e.g., Makishima et al.\ 1986) were similarly
complicated by various observational uncertainties such as mass and
distance, and also by scattering effects (Shimura \& Takahara 1995;
Merloni, Fabian, \& Ross 2000).

At the most basic level, measurements of stellar or disk radii using
blackbody or modified blackbody continua amount to counting photons
under a curve to measure an absolute flux.  This is extremely
difficult (though not impossible), and easily complicated by
additional continuum components and detector flux calibration
uncertainties.  Partially owing to difficulties associated with
continuum spectroscopy, and partially owing to the detection of broad,
possibly relativistic Fe K emission lines in Cygnus X-1 (Barr, White,
\& Page 1985) and 4U 1543$-$475 (van der Woerd, White, \& Kahn 1989;
also see Park et al.\ 2004), parallel efforts to measure fundamental
parameters focused on line spectroscopy.  

Theoretical line models for non-spinning (Schwarzschild; $a = 0$,
where $a = cJ / GM^{2}$) black holes and maximal-spin (Kerr $a =
0.998$; Thorne et al.\ 1974) black holes have been calculated for some
time (Fabian et al.\ 1989; Laor 1991).  Self-consistent models for
hard X-ray illumination of an accretion disk that could give rise to
such emission lines also have a long history (e.g. George \& Fabian
1991).  These models predict that lines arising in the inner disk
should have skewed, asymmetric shapes, including a strong blue wing
that is enhanced by special relativistic beaming and a long red wing
that arises through Doppler shifts and gravitational red-shifts.  Line
shapes are very difficult to discern at the resolution afforded by gas
proportial counter spectrometers (e.g. $E/\Delta E \simeq 6$).
Confirmation of this line shape was first achieved in an {\it ASCA}
observation of the Seyfert-1 galaxy MCG-6-30-15 (Tanaka et al.\ 1995).
It was facilitated by the X-ray charge-coupled devices (CCDs) aboard
the mission, which delivered a resolution of $\sim 12$ in the Fe K
band.  Relativistic iron lines are merely the most prominent part of
the broad-band response of an accretion disk to hard X-rays, known as
the disk reflection spectrum (see, e.g., George \& Fabian 1991;
Magdziarz \& Zdziarski 1995; Dovciak, Karas, \& Yaqoob 2004; Ross \&
Fabian 2005).

X-ray CCDs and relativistic spectroscopy of black holes and neutron
stars are thus intimately linked.  The advanced CCD spectrometers
aboard {\it Chandra}, {\it XMM-Newton}, and {\it Suzaku} have
confirmed relativistic lines in a number of Seyfert-1 spectra (for a
review, see Miller 2007).  Just as importantly, the ability of these
spectrometers to handle high flux levels has made it possible to
clearly detect asymmetric lines in the spectra of stellar-mass black
holes, and to reject the possibility that broad lines are actually a
collection of narrow lines (see, e.g., Miller et al.\ 2001).  If the
inner disk is assumed to be truncated at the innermost stable circular
orbit (ISCO), which is set by the spin of the black hole (see Bardeen,
Press, \& Teukolsky 1972), relativistic lines can be used to infer
black hole spin parameters.  In a number of stellar-mass black holes,
the lines observed are sufficiently broad that varying degrees of
black hole spin may be required (e.g. Miller et al.\ 2002, 2004, 2008;
Reis et al.\ 2009; Hiemstra et al.\ 2010; for a self-consistent
analysis of eight systems, see Miller et al.\ 2009).

Very recently, X-ray CCD spectroscopy of transient and persistent
neutron star X-ray binaries has revealed skewed, asymmetric disk lines
in these systems (see, e.g., Bhattacharyya \& Strohmayer 2007, Cackett
et al.\ 2008, 2009, Di Salvo et al.\ 2009, D'A\`i et al.\ 2009, D'A\`i et
al.\ 2010; for a self-consistent analysis of 10 systems, see Cackett
et al.\ 2010.)  In these sources, relativistic disk lines can be
exploited to constrain the radius of the star, since the stellar
surface (if nothing else) must truncate the disk wherein the lines
arise.  The inner extent of the disk is also related to the Alfven
radius, and so can provide a constraint on stellar magnetic fields.
In the case of the relativistic line observed in the millisecond X-ray
pulsar SAX J1808.4$-$3658 (Cackett et al.\ 2009, Papitto et al.\
2009), the resulting field limits are commensurate with those derived
from X-ray timing (Cackett et al.\ 2009).

New, parallel efforts to derive fundamental properties of compact
objects and inner accretion flows have been fueled by new disk models.
New prescriptions explicitly incorporate inner torque conditions,
radiative transfer through the disk atmosphere, and even black hole
spin parameters (Zimmerman et al.\ 2005, Li et al.\ 2005, Davis \&
Hubeny 2006).  Using these models, black hole spin parameters have
been constrained in a number of systems (see, e.g., McClintock et al.\
2006, Shafee et al.\ 2006; also see Zhang, Cui, \& Chen 1997).  The
improved low-energy range of X-ray CCD detectors (often extending down
to 0.3~keV) relative to current gas detectors (coverage below
$\sim3$~keV is not possible with {\it RXTE} nor {\it INTEGRAL}) allows
for improved measurements of the disk flux, and enables observers to
separate disk emission from line-of-sight absorption in the
interstellar medium (though in many cases dispersive spectroscopy may
be required for this purpose; Miller, Reis, \& Cackett 2009).  Thus,
while the improved energy resolution of X-ray CCD spectrometers is not
necessarily critical for continuum spectroscopy, their energy range
and spectral resolution are beneficial.

While measuring the width of lines is often easier than measuring an
absolute flux, it still requires an accurate characterization of the
underlying continuum flux, and an accurate knowledge of how the
spectrometer reacts to high flux levels.  The latter issue is
typically unimportant for gas proportional counter spectrometers, but
it can be important for X-ray CCD spectrometers.  Indeed, the reaction
of X-ray CCDs is important for both forms of relativistic spectroscopy
of compact objects: an instrumental failure to accurately record the
flux level is problematic for disk continua, and distortions to the
continuum shape and energy response are problematic for line
spectroscopy.  Photon pile-up occurs when two or more photons land
within a detection cell within a single CCD frame time.  This causes a
degeneracy between detecting ${\rm N}$ photons with energy ${\rm
E}_{i}$ and a single photon with energy of $\Sigma_{\rm i}^{\rm
N}~{\rm E}_{\rm i}$.  Thus when pile-up occurs, energy and flux
information are lost.

It has recently been suggested that ineffective pile-up mitigation
could produce spurious results concerning black hole spin, neutron
star radii, and the radial extent of accretion disks as a function of
the mass accretion rate (e.g. Yamada et al.\ 2009; Done \& Diaz Trigo
2010, Ng et al.\ 2010).  Advanced statistical descriptions of photon
pile-up in X-ray CCD detectors have been developed (Davis 2001; also
see Ballet 1999), and implemented into fitting packages such as XSPEC
and ISIS.  The Davis et al.\ (2001) model has been applied in a number
of regimes to correct for photon pile-up distortions, including
spectra from: isolated neutron stars (e.g. van Kerkwijk et al.\ 2004),
ultra-luminous X-ray sources (e.g. Roberts et al.\ 2004), transient
Galactic black hole candidates (e.g. Jonker et al.\ 2004), low-mass
X-ray binaries (e.g. Heinke et al.\ 2006), and even AGN (e.g. Wang et
al.\ 2010).  Most notably, Nowak et al.\ (2008) employed the basic
elements of the Davis model to correct distortions to the {\it
Chandra} spectra of the black hole candidate 4U 1957$+$11.  Inner disk
parameters consistent with those derived using spectra from other
missions, including {\it XMM-Newton} and {\it RXTE}, were then
recovered.  A similar treatment of {\it Chandra} spectra of Cygnus X-1
by Hanke et al.\ (2009) was able to bring different spectra into close
agreement.

With the aim of developing a rigorous understanding of how pile-up
may affect relativistic spectroscopy, we have conducted extensive
simulations based on a range of assumed input spectra, flux levels,
and detector properties.  This effort is timely as X-ray CCD
spectrometers will be the standard in the field for the foreseeable
future, at least for moderate resolution spectroscopy of bright
targets.  In the following sections, we briefly review the operation
of X-ray CCD detectors, review the pile-up model used in our
simulations, describe our simulations and results, and discuss the
impacts of our findings.

The primary goal of this analysis is not to understand how to best
avoid photon pile-up, though the results offer some insights on this
point.  Rather, the goal of this exercise is to understand the nature
and magnitude of distortions to relativistic disk parameters when
photon pile-up is present but unknown to an observer, poorly
quantified, or difficult to mitigate in a rigorous manner.  In a
subset of our simulated spectra, then, photon pile-up is quite severe.
When faced with similar real data, an observer would likely be
motivated to extract an annulus or to otherwise at least partially
mitigate the effects of pile-up.  

\section{The Basics of X-ray CCD Spectrometers}

At the most basic level, a CCD is an array of coupled capacitors.  When
an incident photon interacts in the semiconductor substrate, electrons
are liberated.  The charge is collected and stored in pixels, and it
can be transfered to neighboring pixels -- and eventually into an
amplifier and read-out system -- by virtue of the coupling.  

In many respects, the operation of X-ray CCDs is similar to that of
optical CCDs.  One important difference is that optical photons only
liberate a small number of electrons (typically zero or one) in an
interaction.  X-ray photons have much more energy, of course, and
liberate many electrons.  Indeed, the number of electrons liberated in
an interaction is proportional to the energy of the incident X-ray.
This fact means that X-ray CCDs are not merely imaging instruments,
but medium-resolution spectrometers as well.  The gas proportional
counters aboard {\it RXTE} have an energy resolution of $\sim6$;
current X-ray CCDs (e.g. {\it Chandra}/ACIS) achieve an energy
resolution of $\sim30$ at 5.9 keV.

Most X-ray CCDs cover the 0.3--10.0~keV band.  These limits are set by
a combination of factors, including the efficiency of capturing and
clocking charge clouds generated by low energy X-rays (this is
partially set by the electrodes), the depletion layer thickness, the
ability of current X-ray mirrors to focus hard X-rays, and the ability
of the instrument to distinguish hard X-rays from non-X-ray events.
The detailed performance characteristics of a given X-ray CCD depend
on several variables, including the nature of the device structure (n-
or p- substrate, MOS or pn junction); whether charge is stored and
read-out close to the X-ray-illuminated face of the CCD, or far from
the illuminated face; the temperature at which the CCD is being
operated (this affects dark currents and noise), and a number of more
subtle factors.

Photons are detected and elementary screening of desirable
(e.g. X-ray) versus undesirable (e.g. cosmic ray) events is achieved
in part by assigning event ``grades''.  Interactions are characterized
using event ``boxes'', which are usually 3x3 cells of pixels.  The
pattern of pulseheights that are consistent with photon interactions
can be defined and differentiated from ``hot'' pixel and cosmic ray
patterns.

Efficient spectroscopy with an X-ray CCD depends on each event box
recording zero or one photons per CCD frame time.  When the X-ray flux
incident on a CCD is too high, multiple photons can be read as a
single high energy photon (see above).  The resulting flux is thus
registered as falsely low, and the resulting spectrum as falsely hard.
Grade migration provides a potential means of recognizing and diagnosing the
degree of photon pile-up suffered in a given observation.  When
pile-up is important, charge patterns consistent with single photon
interactions are reduced and patterns consistent with higher energy
photon interactions or multiple photon interactions are increased.

Various strategies can be deployed to deliver nominal spectroscopic
response even at high X-ray flux levels.  The nominal CCD frame time
can be greatly reduced, thus lowering the probability of two photons
registering in a given event box.  This often comes at the expense of
imaging information, observing efficiency, or both.  Over-sampling a
broad telescope PSF with a high number of event boxes is another
viable means of limiting the number of photons that may land in a
given event box.  The expense of a broad PSF is, of course, the loss of
fine image quality.  Observers can try to further prevent or limit
photon pile-up by only extracting events from the wings of the PSF in
annuli, by placing a source off of the optical axis of the telescope
to blur its flux over more pixels, or by adopting a rigid grade
selection that permits only single-photon events.  In practice,
however, some spectra obtained with X-ray CCDs will suffer from photon
pile-up.

\section{The Davis Photon Pile-up Model}

Pile-up is a complex but well-understood statistical process: At a
given flux level, a given event box has a fixed probability of getting
0, 1, or N photons.  The probability of a given event grade for each
outcome is also well-determined.  If the input spectrum is known, or
if it can be estimated, the distribution of ``good'' event grades
(those likely to originate from actual X-ray photons) can be used to
infer the extent and severity of photon pile-up.  On this basis, Davis
(2001) has developed a model of photon pile-up.  The model not only
accounts for the energy shifts due to photon pile-up, but also for
flux decrements due to grade migration.

The crux of the Davis model is a new integral equation for the number
of counts detected from a point source.  The conventional equation for
the number of counts detected in an X-ray spectrometer is linear, and
discussed in detail in Gorenstein, Gursky, \& Garmire (1968).
However, pile-up is a nonlinear process, and requires a more advanced
treatment.  Davis (2001) presents an alternative integral equation
that accurately characterizes the counts observed in the presence of
photon pile-up.  A fundamental assumption of the pile-up model is that
a charge cloud in the CCD arising from N photons can be treated as the
linear superposition of N individual charge clouds.  Since the drift
time for charge clouds in a CCD is typically much less than a
microsecond, which is much smaller than typical readout times, and
since a CCD amplifier is highly linear at X-ray signal levels, this
assumption is valid for all of the CCD modes that will be considered
in this work.

In practice, the chief uncertainty in correcting for pile-up using
this model is that the details of the input spectrum are usually not
known a priori.  Exceptions might include instances where different
missions are simultaneously observing the same source, or an instance
wherein a given CCD spectrometer alternates between different
operational modes.  {\it This uncertainty does not enter into our
  analysis.}  Simulating the effects of photon pile-up by convolving
different input spectra with the pile-up model of Davis (2001)
is a well-controlled experiment -- the nature of the input spectrum is
known perfectly.  The results of this procedure can be thought of as
characterizing the nature and severity of systematic errors on
relativistic disk parameters arising due to photon pile-up.

It is worth noting that although the pile-up model was developed for
{\it Chandra}, it is easily applicable to other missions.  The crucial
modification in describing pile-up in other CCD spectrometers is to
accurately account for the fraction of the telesope point spread
function (PSF) that is enclosed in each event box.  This can differ
considerably from mission to mission.  Whereas a single event box
captures a high fraction of {\it Chandra's} PSF, a single event box
captures a very small fraction of {\it Suzaku's} PSF.  In the latter
case, to evaluate pile-up in a reasonable extraction region, many
event boxes must be considered jointly.

As implemented in the XSPEC spectral fitting package (Arnaud \& Dorman
2000), the Davis pile-up model has six parameters: the CCD frame time,
the maximum number of photons to pile-up, the grade correction for a
single photon, the grade morphing parameter (the grade migration
function is a probability assumed to be proportional to
$\alpha^{p-1}$, $0 \leq \alpha \leq 1$, where $\alpha$ is the grade
morphing parameter and $p$ is the number of piled photons), the PSF
fraction considered, and the number of event regions.  This
description of grade migration is based on the premise that photon
arrival times in a given cell obey Poission statistics.  It correctly
captures the fact that the probability of recording an unwanted event
grade increases with the number of piled photons.

\section{The Simulated Spectra}
To understand the specific effects of photon pile-up on disk continua
and reflection parameters, we created a large number (325) of
simulated spectra using the Davis pile-up model.  Realistic,
multi-component spectral forms based on published results were
employed to make the simulations as realistic as possible.  The
spectra were generated using the ``fakeit'' command in XSPEC version
12.  For simplicity, and because this exercise is necessarily
concerned with bright sources, no backgrounds were used in making the
source spectra.  In all cases, an exposure time of 100,000 seconds was
used when creating simulated spectra.  This exposure is longer than a
typical observation, but allows for excellent photon statistics.

In all of the simulations made in this work, default values were
assumed for the maximum number of photons to pile-up (5) and the grade
correction for a single photon (1).  The frame time, PSF fraction, and
number of event boxes were set according to the telescope and CCD
combination used in each simulation.  With these parameters fixed, the
grade morphing parameter $\alpha$ effectively sets the severity of
photon pile-up (when the incident source flux in each event box leads
to pile-up).    

As implemented in XSPEC, the pile-up model is a convolution model.
Input spectra were convolved with the pile-up kernel to produce a
resultant spectrum that is distorted by photon pile-up effects.  In
practice, for a given detector and mode, the degree of pile-up is set
by the incident flux level.  In our simulations, the degree of photon
pile-up was controlled by adjusting the grade morphing parameter
$\alpha$.  Steps of 0.1 were used for $0.1 \leq \alpha \leq 0.8$, and
steps of 0.04 were used for $0.8 \leq \alpha \leq 0.99$.  The step
size was reduced for high values of $\alpha$ in order to provide
better resolution when pile-up typically starts to introduce strong
distortions.  For each combination of input spectrum and detector
mode, then, 13 simulated spectra were generated.

\section{Input Spectral Forms}

X-ray binaries display a wide variety of phenomena.  Periods of
correlated multi-wavelength behaviors can be classified into
``states'' that may correspond to distinct changes in the accetion
flow (for a review, see, e.g. Remillard \& McClintock 2006; also see
Belloni et al.\ 2005).  In this work, we consider three
commonly-recognized ``states'' for black hole X-ray binaries, and a
single input spectral form for neutron star low-mass X-ray binaries
(the so-called ``Z'' and ``atoll'' sources).  In the interest of
simplicity and reproduciblity, we have adopted phenomenological
spectral models for each state with values consistent with those
reported in the literature.  Table 1 lists values for all of the input
spectra considered in this work.  The paragraphs below offer some
context and motivation for why certain models and values were used to
generate the spectra.

{\it The Very High State}: This state can be extremely bright, and it
has seldom been observed using CCD spectrometers.  Continua observed
in this state often consist of a hot disk and steep power-law
component.  We assumed the continuum parameters measured in the very
high state of GX 339$-$4 using the {\it XMM-Newton}/EPIC-pn in
``burst'' mode (Miller et al.\ 2004).  This mode has a frametime of
just 7 $\mu$sec and a deadtime of 97\%, but it is effective at
preventing photon pile-up for fluxes up to 6~Crab.  The particular
spectrum chosen as a template is broadly consistent with very high
state spectra obtained from {\it RXTE} monitoring of other transients
(e.g. 4U 1543$-$475, Park et al.\ 2004; GRO J1655$-$40, Sobczak et
al.\ 1999; XTE J1550$-$564, Sobczak et al.\ 2000, Miller et al.\
2003).

The ``intermediate'' state is similar to the ``very high'' state in
terms of its spectral form (both thermal disk and non-thermal
power-law-like emission are important), but typically has a lower
flux.  The effects of photon pile-up in this state are likely to be
similar to the effects on very high state spectra at low values of the
grade migration parameter $\alpha$.  The ``intermediate'' state is not
treated separately in this work.

{\it The High/Soft State}: This state can also be quite bright, but it
typically persists for a longer period than the ``very high state''.
As a result, it is more commonly observed with CCD spectrometers.
Continua observed in this state are strongly dominated by a hot disk
component, accompanied by a weak, steep power-law.  To model the
high/soft state, we selected one of the spectrally softest
observations of 4U 1543$-$475 obtained using {\it RXTE} (Park et
al.\ 2004).  This observation occurred on MJD 52475, and is consistent
with the spectra selected for relativistic disk spectroscopy by Shafee
et al.\ (2006).  Here again, this high/soft state is typical of
spectra of other black hole transients in the same state (see, e.g.,
Sobczak et al.\ 1999, 2000).

{\it The Low/Hard State}: The low/hard state has been a frequent
target with CCD spectrometers recently, in an effort to understand the
accretion flow at low mass accretion rates (for a recent survey, see
Reis, Fabian, \& Miller 2010, also see Tomisck et al.\ 2009).
Spectral continua in this state are dominated by a hard power-law,
sometimes with weak emission from the accretion disk.  The continuum
spectrum assumed in our simulations is a composite of the parameters
reported in observations of GX~339$-$4 by Miller et al.\ (2006),
Tomsick et al.\ (2008), and Wilkinson \& Uttley (2009).

{\it A Typical Z/Atoll Spectrum}: Unlike most black hole X-ray
binaries with low-mass companions, neutron stars with low magnetic
fields are typically persistent (but variable) sources.  Whereas black
hole continua can often be well-characterized in terms of thermal disk
emission and a power-law, neutron star spectra often require a third
component (Lin, Remillard, \& Homan 2007; Cackett et al.\ 2008,
Cackett et al.\ 2010, D'A\`i et al.\ 2010).  A simple blackbody likely
corresponds to emission from the boundary layer between the accretion
disk and stellar surface (Revnivtsev \& Gilfanov 2006).  Emission from
this region may be Comptonized, but blackbody emission that is
Compton-upscattered in a region with high $\tau$ and low $kT_{e}$ can
be modeled with a hotter, smaller blackbody (London, Taam, \& Howard
1986).  We used the spectral continuum parameters reported by Cackett
et al.\ (2008) for Serpens X-1.

\subsection{Relativistic Line Properties}

Relativistic line components were added to all continuum spectra,
apart from those for the ``high/soft'' state.  The ``Laor''
relativistic emission line model (Laor 1991) was assumed to simulate emission
lines from the inner disk.  In all cases, the input inner radius was
fixed at $6~GM/c^{2}$, the emissivity index was fixed at $q=3$ (where
$J(r) \propto r^{-q}$), and the inclination was fixed at $30^{\circ}$.
This radius is substantially larger than the innermost stable circular
orbit for a maximally-spinning Kerr black hole ($1.24~GM/c^{2}$), and
significantly larger than measured in a number of stellar-mass black
holes (Miller et al.\ 2009).  However, one purpose of this
investigation is to see if photon pile-up can artificially {\it
  broaden} relativistic lines, giving falsely small inner radii and
falsely high spin values.  Setting the inner radius at $6~GM/c^{2}$,
then, permits distortions due to photon pile-up ample opportunity to
produce falsely broad lines.  For all simulated black hole spectra,
the input flux of the relativistic ``Laor'' line was normalized to
give an equivalent width of $300$~eV as per Miller et al.\ (2004).

Similarly, $6~GM/c^{2}$ is approximately 12.4~km for a 1.4~$M_{\odot}$
neutron star -- only slightly larger than the canonical stellar radius
of 10~km employed in many circumstances.  Assuming the stellar surface
(if not the boundary layer) must truncate the accretion disk, a
relativistic line suggesting a smaller radius might be taken as
evidence of a ``strange'' star.  In the case of neutron stars, then,
fixing the input radius at $6~GM/c^{2}$ serves to assess the ability
of photon pile-up to give false evidence of ``strange'' stars through
falsely small disk radii.  For all simulated neutron star spectra, the
input flux of the ``Laor'' line was normalized to given an equivalent
width of $150$~eV as per Cackett et al.\ (2008).

Iron emission lines in accreting black holes and neutron stars are
merely the most prominent part of the disk reflection spectrum
(e.g. Dovciak, Karas, \& Yaqoob 2004; Ross \& Fabian 2005).
Self-consistent modeling of real spectra requires modeling of the
entire reflection spectrum, not just the broadened emission line.  For
simplicity, and because the line drives constraints on inner disk
radii (and therefore spin) in fits to real data, disk reflection
models are not treated in this analysis.

\subsection{The Thermal Disk Continuum}

In all spectra, thermal continuum emission from the accretion disk was
simulated using the simple ``diskbb'' model (Mitsuda et al.\ 1984)
within XSPEC.  Using this model, the inner disk radius can be derived
via: $r = (d / 10 {\rm kpc}) \times (K/{\rm cos}\theta)^{1/2}~{\rm km}$,
where $K$ is the model flux normalization, $d$ is the distance to the
source, and $\theta$ is the inclination of the inner disk.  If the
mass of the source is known, this radius can be converted to
gravitational units.  

This is merely a ``color'' radius, however, and does not account for
the effects of spectral hardening due to radiative transfer through
the disk atmosphere (Shimura \& Takahara 1995; Merloni, Fabian, \& Ross 2000).
It is well-known that this model is overly simple in other ways.  For
instance, it does not include a zero-torque condition at the inner
edge, whereas including this condition can reduce implied radii by a
factor of $\sim$2 (Zimmerman et al.\ 2005).  Moreover, a number of new
disk models have been developed in which black hole spin is an
explicit parameter (Davis \& Hubeny 2006).

Deriving spins using these newer, more physical models requires
extremely accurate knowledge of the absolute disk flux, however, which
can be complicated by a number of effects (e.g. the flux calibration
of the detector, the accuracy to which line of sight absorption is
known, the nature of the hard component, etc.).  And whereas new
models can have as many as 10 free parameters, the ``diskbb'' model is
able to accurately characterize the thermal continuum with only two
(temperature and flux normalization).  Owing to its simplicity and
prior application to many neutron star and black hole spectra over
numerous years and X-ray missions, then, all thermal disk continua
were simulated using the ``diskbb'' model, and our analysis is
primarily concerned with changes in the flux normalization parameter.

In the Z/atoll spectra, a simple (single-temperature) blackbody
function was used to describe emission from the stellar surface,
independent of the thermal emission from the disk.

\subsection{Interstellar Absorption}

Each input spectrum was modified by interstellar absorption using the
``tbabs'' model (Wilms, Allen, \& McCray 2000).  In all cases, the
equivalent neutral hydrogen column density was fixed at ${\rm N}_{\rm
  H} = 5.0 \times 10^{21}~ {\rm cm}^{-2}$.  This is a moderate value,
consistent with columns that have facilitated studies of the thermal
disk continuum in a number of sources and at different mass accretion
rates.  

\subsection{Simulations with Gaussian Lines}

Although a number observations made with the {\it Chandra}/HETGS
(e.g. Di Salvo et al.\ 2005) and with {\it Suzaku} (Cackett et al.\
2008, Reis, Fabian, \& Young 2009, Cackett et al.\ 2010) find broad,
relativistic line shapes in the spectra of neutron stars, such lines
are less established in neutron star spectra than in black hole
spectra.  Ng et al.\ (2010) suggest that neutron star lines may
actually be narrower ($\sigma = 0.33$~keV, on average) and symmetric.
Therefore, we also simulated neutron star spectra using the same
neutron star continuum described above and in Table 1, but with
Gaussian model with ${\rm E} = 6.7$~keV, $\sigma = 0.33$~keV, and an
equivalent width of 150~eV, rather than a relativistic line.

\section{Detectors and Modes}
The CCD spectrometers and modes considered in our simulations are
detailed in Table 2.  The modes selected reflect those that users
might realistically select when observing bright sources.  The
``burst'' mode of the EPIC-pn camera aboard {\it XMM-Newton} is not
considered, as it is capable of handling exceptionally high flux
levels without pile-up distortions.  Common imaging modes available on
the ACIS and XRT instruments aboard {\it Chandra} and {\it Swift},
respectively, are also not considered; these are known to readily
suffer pile-up for even moderately bright sources and are typically
avoided by users.  Finally, with {\it Chandra}, pile-up can be largely
mitigated while obtaining a high-resolution spectrum with the HETGS;
however, assessing pile-up in this circumstance is especially complex
and beyond the scope of this investigation.

An ``effective'' frame time is given in Table 2.  This is the frame
time that is important to consider when assessing the extent and
impacts of photon pile-up.  It multiplies the time required to extract
one element of charge (typically a pixel, but sometimes a larger
``macropixel'') by the number of elements necessary to define an event
box.  (Some missions -- but not all -- redefine event boxes for
different modes; see the discussion below.)  If one could know a
priori that a sequence of events were all ``singles'' wherein charge
from photon events was contained in only one pixel, then the effective
frame time would equal the time required to clock just one row of
pixels.  However, it is not possible to know this a priori, and {\it
  in practice one only knows that an event was a ``single'' if the
  other rows in an event box are free of charge}.  Thus, for pile-up,
the time scale of concern is the time required to transfer a full
event box.

A significant simplification in our simulations is that the grade
morphing parameter $\alpha$ is assumed to be constant in each
extraction region.  In practice, this is not true; $\alpha$ would
depend on radius as an actual telescope PSF focuses successively less
energy into successively larger annuli.  Our simulations therefore
capture the strongest impacts of photon pile-up distortions, and
ignore how less distorted spectra from large annuli might affect
results when summed with spectra from smaller annuli.  In a partial
effort to minimize the impact of the assumption of a constant
$\alpha$, the extraction regions assumed in our simulations do not
encircle the same fraction of the total incident energy, but rather
attempt to sample the part of the PSF where the encircled energy
changes only linearly with radius.  This is not practical in the case
of {\it Chandra}, however, which has a very tight PSF.  A secondary
simplification is that the pile-up model and our simulations do not
consider how charge from event boxes might bleed into adjacent boxes
in the case of severe pile-up.

Additional details concerning the individual spectrometers and modes
considered, and how they are treated in our simulations, are given below:

\subsection{Chandra/ACIS Continuous Clocking Mode}

ACIS event boxes consist of a 3x3 box of pixels, each approximately
0.5'' on a side.  In continuous clocking or ``CC'' mode, charge is
continuously transferred from the CCD (usually ACIS-S3) at a rate of
2.85~msec per row.  Two-dimensional imaging is sacrificed in favor of
fast read-out.  The definition of ACIS event boxes is unchanged
between standard imaging modes and CC mode.  In our simulations, then,
we have assumed an effective frametime of 8.55~msec, since this is the
time require to clock a full event box.

As noted above, given the tight PSF of {\it Chandra}, it is
impractical to extract photons from a region that encircles less than
90\% of the incident energy.  In our simulations, we therefore assumed
an extraction radius of 2.5'', which corresponds to 90\% of the
incident energy.  In the one-dimensional regime of CC mode, the
extraction region is then a bar 5'' in length.  Note that this means
that only 3.3 event boxes tile the extraction region.

Where possible, generic, readily-available response matrices were
used in order to facilitate checks on this work by other teams.  To
simulate ACIS CC mode spectra, then, we used the ACIS response
matrices made available by the mission for simulations supporting
Cycle 11 proposals: ``aciss\_aimpt\_cy11.rmf'' and
``aciss\_aimpt\_cy11.arf''.  For more information about the ACIS
spectrometer and CC mode, please see Garmire et al.\ (2003) and the
{\it Chandra} Proper's Observatory Guide
(http://cxc.harvard.edu/proposer/POG/).

\subsection{Swift XRT Windowed Timing Mode}

In ``windowed timing'' or ``WT'' mode, charge rows are grouped by 10
in the read-out direction (effectively creating ``macropixels'') and
rapidly tranferred from the CCD.  In the standard photon counting
mode, a 3x3 grouping of pixels (each pixel is 2.36'' on a side) is
used to define an event box.  However, in windowed timing mode, the
event box is altered to be 10x7 pixels: 1 macropixel in the clocking
direction, and 7 pixels in the orthogonal direction.  This change
improves the ability of the detector to identify and screen events in
windowed timing mode.  Note that because the new event box is only 1
macropixel in the clocking direction, the nominal and effective frame
times are the same in windowed timing mode (see Table 2).  Like {\it
Chandra}/ACIS ``continuous clocking'' mode, ``windowed timing'' mode
has a livetime fraction of 1.0.

In our simulations, we assumed an extraction radius of 9.0'', which
roughly corresponds to the half power radius for the {\it Swift}/XRT.
The extraction region is tiled by only 1.1 event boxes.  For
simplicity, we have also assumed an effective frame time of 1.77~msec
(see Table 2).  In practice, depending on where a photon strikes, it
may take up to two read-out times of 1.77~msec to clock the charge
through the extraction region (which roughly matches the half-power
diameter of the PSF).  In this sense, our simulations slightly
under-estimate the severity of pile-up distortions to {\it Swift}/XRT
spectra obtained in ``windowed timing'' mode.

To simulate ``windowed timing'' mode spectra, we used current,
standardized response functions available through the HEASARC
calibration database, ``swxwt0to2s6\_20070901v011.rmf'' and
``swxwt0to2s6\_200101 01v011.arf''.  These responses have been
developed specifically for WT mode, and differ from the responses
appropriate for normal imaging modes.

For more information about the {\it Swift}/XRT, please see Hill et
al.\ (2004), Burrows et al.\ (2005), the {\it Swift} Technical
Handbook
(http://swift.gsfc.nasa.gov/docs/swift/proposals/appendix\_f. html) and
the XRT User's Guide
(http://swift.gsfc.nasa.gov/docs/ swift/analysis/xrt\_swguide\_v1.2.pdf).

\subsection{XMM-Newton EPIC-pn ``Timing'' Mode}

The ``timing'' mode of the EPIC-pn camera is similar to the ``windowed
timing'' mode of the {\it Swift}/XRT in some respects.  To achieve a
high time resolution, normal CCD pixels are grouped into
``macropixels''.  Each macropixel is 10 pixels in the read-out
direction and 1 pixel in the orthogonal direction.  Since each pixel
is 4.1'' on a side, this means that a single macropixel is actually
41'' in the read-out direction.  In its long dimension, then, a single
macropixel is equivalent to the 70\% encircled energy diameter of the
telescope.  In standard imaging modes, {\it XMM-Newton} event boxes
are the standard 3x3 pixel element, as per other X-ray CCD cameras.
In timing mode, however, an event box is 3 macropixels by 3
macropixels.  This means that a single event box is 123'' long
(equivalent to the 95\% encircled energy diameter) and 12.3'' wide
(appoximately equal to the half-power diameter).  This event box is
extremely large, and roughly equivalent to the extraction region that
is often used by observers using this mode.  

Creating macropixels enables a short read-out time of 0.03~msec {\it
  per macropixel}.  This translates to an effective frame time of 0.09
  msec {\it per event box}.  Again, this is a simplification that may
  serve to under-estimate the severity of photon pile-up distortions
  to ``timing'' mode spectra.  In practice, each macropixel takes at
  least three read-out times to be completely clocked, and it may take
  four timescales depending on where the photon strikes.  Thus, each
  macropixel may contain more charge than anticipated in these simplified
  simulations.

The short frame time achieved in ``timing'' mode is impressive,
especially considering that ``timing'' mode operates with a live-time
fraction of 0.99.  Unlike its cousin, ``burst'' mode (live-time
fraction: 0.03), ``timing'' does not handle high fluxes by making
exceptionally short exposures while discarding the bulk of the
incident flux.  These abilities come at the expense of tiling the PSF
with many event boxes, however, which is another important means of
reducing the severity of photon pile-up distortions.  

A distinctive feature of the {\it XMM-Newton}/EPIC-pn camera is that
diagonally adjacent pixels are treated as two single pixel events.
Differences between detectors at this level are assumed to be small.

Generic and separate redistribution and ancillary matrix files are not
readily available to {\it XMM-Newton} users.  Timing mode response
matrices were generated using the ``rmfgen'' and ``arfgen'' tools in
SAS version 9.0.0.  These matrices where then used to simulate all
timing mode spectra.  

For more information about {\it XMM-Newton} and the EPIC-pn camera,
please see Str{\"u}der et al.\ (2001), Haberl et al.\ (2004), and the
User's Handbook
(http://xmm.esac.esa.int/external/xmm\_user\_support/ documentation/index.shtml).

\subsection{XMM-Newton EPIC MOS ``Full Frame'' Mode}

``Full frame'' mode is a standard imaging mode for the MOS 1 and MOS 2
cameras aboard {\it XMM-Newton}, not a specialized timing mode like
those described above.  There are two reasons for considering it in
this analysis.  First, the small, 1.1'' pixels of the MOS cameras
allow for a large number of event boxes to tile an extraction region,
which is one viable strategy for reducing the effects of photon
pile-up.  Second, by extracting annuli that excise central regions
within a piled-up source image, observers are often able to make use
of data obtained in this and similar operating modes.  Whether or not
pile-up is totally mitigated through this procedure is sometimes in
doubt, and this analysis can help to clarify the nature of any
distortions imposed on the extracted spectrum by residual pile-up.

In simulating MOS ``full frame'' spectra, circular extraction regions
with 20'' radii were assumed.  Within this range, the encircled energy
fraction of the PSF changes almost linearly with radius.  The regions
assumed would encircle approximately 70\% of the total incident
energy.  As with pn ``timing'' mode, responses were generated using
the ``rmfgen'' and ``arfgen'' functions within SAS version 9.0.  These
responses were then used to create all simulated ``full frame''
spectra.  

For additional information on the {\it XMM-Newton}/EPIC-MOS
cameras, please see Turner et al.\ (2001).

\subsection{Suzaku Windowed Burst Mode}

The XIS spectrometer aboard {\it Suzaku} provides many different
operational modes that can be used to optimize science returns and to
reduce or eliminate photon pile-up.  Each XIS CCD has a nominal frame
time of 8.0 seconds, but this can be reduced by choosing a ``window''
option.  For instance, using a 1/4 window means that only 1/4 of the
CCD will be exposed, reducing the frame time to 2.0 seconds.  (Smaller
windows are possible but cannot be used owing to drift in the
spacecraft pointing over the course of an observation.)  The frame
time can be further reduced by selecting a ``burst'' option; in this
case, the CCD only exposes for a fraction of the nominal frame
time.  The remainder of the frame time is then dead time, reducing the
overall observing efficiency and requiring longer exposures.  For this
analysis, the second-most conservative combination of window and burst
option has been considered.  For all simulated spectra, we assumed a
1/4 window and 0.3 second burst option.

It should be noted that the XIS can be operated in ``PSUM'' mode,
wherein the full height of the CCD is read-out in the time normally
required to clock one row of charge.  In some respects, this mode is
similar to the ``timing'' and ``burst'' modes available on the {\it
XMM-Newton}/EPIC-pn camera, and for similar reasons it is actually
less effective at mitigating pile-up than the most conservative
combinations of XIS window and burst options.

Each simulated spectrum assumed a circular extraction region with a
radius of 60'', which encircles approximately half of the incident
energy for the {\it Suzaku} PSF.  Within this radius, the encircled
energy fraction changes in a roughly linear way with radius.  Each XIS
pixel is 1'' on a side, and standard 3 pixel by 3 pixel boxes are used
to define event grades.  Within the extraction region, then, there are
1256.0 event boxes.  Tiling a broad PSF with many event boxes acts to
reduce the flux per event box.  

For more information on the
XIS, please see the {\it Suzaku} ABC Guide
(http://heasarc.gsfc.nasa.gov/docs/suzaku/analysis/abc/) and the {\it
Suzaku} Technical Description
(http://heasarc.gsfc.nasa. gov/docs/suzaku/aehp\_prop\_tools.html).

\section{Analysis and Results}
%
% \subsection{Spectral Fitting Methodology}
%
All simulated spectra were fit using the same model components that
were employed to generate the spectrum, but without using the pile-up
kernel.  This procedure permits an examination of how spectral fitting
results are distorted by photon pile-up effects, when such effects
have not been mitigated entirely (e.g. by extracting events in an
annulus, by selecting a shorter frame time, etc.).  For the purpose of
illustrating how parameters are distorted even in the event of severe
photon pile-up, fits were made even when the $\chi^{2}$
goodness-of-fit statistic was not acceptable.  All errors reported in
this analysis are $1\sigma$ confidence errors, and were determined
using the ``error'' command within XSPEC.  In the cases where the fits
were unacceptable, the errors are not true $1\sigma$ errors.

A few simplifying assumptions were made in fitting the simulated
spectra.  The equivalent neutral hydrogen column density was held
fixed at the input value, ${\rm N}_{\rm H} = 5.0\times 10^{21}~ {\rm
cm}^{-2}$.  In practice, this might be equivalent to assuming the
column density measured via a sensitive {\it Chandra}/HETGS or {\it
XMM-Newton}/RGS spectrum in which individual neutral photoelectric
absorption edges were detected (e.g. Miller, Cackett, \& Reis 2010).
Some of the relativistic iron line parameters were constrained to
ranges and/or fractional uncertainties that are common in the
literature.  Specifically, all fits to the simulated line spectra
assumed $6.70~ {\rm keV} \leq {\rm E} \leq 6.97~ {\rm keV}$, $3 \leq q
\leq 5$, and $10^{\circ} \leq \theta \leq 50^{\circ}$.  Fits to the
simulated spectra were otherwise unconstrained; line and continuum
model parameters were free to take on whatever value best described
the distorted spectra.

The goal of this analysis is simply to understand how disk reflection
and disk continua are affected by pile-up.  A detailed treatment of
distortions to (more physical) continuum spectra (e.g. Comptonization
spectra, and/or coupled disk plus corona models such as ``eqpair'')
might also be timely, but it is beyond the scope of this paper.  The
sections that follow are narrowly focused on the results of fits to
relativistic lines and simple disk continua.

\subsection{Pile-up and Relativistic Lines}

The results of these limited exercises are clear: when photon pile-up is
severe, it causes relativistic emission lines in black hole spectra to
be measured as falsely narrow.  In fits to the simulated spectra with
the Laor line model, measured radii were all consistent with or
significantly larger than the true value of $6~{\rm GM}/{\rm c}^{2}$.
Figures 1 and 2 trace the evolution of measured radius as a function
of pile-up severity (set by the grade migration parameter $\alpha$).
Figure 3 shows how the entire spectrum is affected by pile-up.  Figure
4 shows a sequence of line profiles from simulations with increasing
photon pile-up.  The same results hold in the case of our simulated
neutron star spectra, with one modest exception (see below).

Even the baseline flux level assumed in our ``very high'' state
simulations (corresponding to low values of $\alpha$) is found to
completely overwhelm {\it Chandra}/ACIS ``continuous clocking'' mode
and {\it XMM-Newton}/EPIC-pn ``timing'' mode.  The emission line could
not be detected nor reliably fit except at low values of the grade
migration parameter $\alpha$.  In any real observation at an even
higher flux level (corresponding to a higher value of $\alpha$), then,
a line may not be clearly detected.  Similarly, for the {\it
XMM-Newton}/EPIC MOS in ``full frame'' mode, event loss dominates over
grade migration in the simulated spectra, and so the {\it shape} of
the spectrum -- including the line -- is preserved though the line and
continuum flux are a small fraction of their input values.  The
results shown in Figure 1 do not indicate that the MOS ``full frame''
mode is equipped to deal with typical ``very high'' state flux levels.

The {\it Swift}/XRT in ``WT'' mode and {\it Suzaku}/XIS (with 1/4
window and 0.3~second burst option) are less overwhelmed by the high
flux associated with our baseline ``very high'' state spectrum.  In
fits to those simulated spectra, the inner disk radius correlates with
the severity of pile-up as traced by the grade migration parameter
$\alpha$ (again, $\alpha$ serves as a proxy for increasing the flux
over the baseline levels given in Table 1).  The results shown in
Figure 1 indicate that the {\it Suzaku}/XIS is actually best-equipped
to deal with such high flux levels, although even in this case fits to
the iron line fail to recover the full line width and give falsely
large inner disk radii.  The apparent fluctuations in inner disk
radius at high values of $\alpha$ seen in the XRT trend in Figure 1
are likely the result of errors that are under-estimated because the
overall fit is poor.

In the ``low/hard'' state, the degree of spectral distortions is
generally reduced with respect to the ``very high'' state (see Figure
2).  The most accurate inner disk radii were recovered in the
simulated {\it Suzaku}/XIS spectra.  A 1/4 window option and 0.3
seconds burst option appear to suffer minimal disortion due to photon
pile-up, at least for the flux levels that would generate the values
of $\alpha$ that were simulated.  Photon pile-up distortions are more
pronounced in the other spectra: measured inner disk radii are found
to depart more strongly from the true value with increasing $\alpha$.
The {\it XMM-Newton}/EPIC-pn ``timing'' mode and {\it Swift}/XRT
``windowed timing'' mode appear to only suffer modest distortions as
pile-up becomes more severe: in each case, lines are measured to be
falsely narrow and to give inner disk radii that are too large by
2--3~${\rm GM}/{\rm c}^{2}$.  Fits to the simulated {\it
XMM-Newton}/EPIC MOS spectra in ``full frame'' mode show the most
marked trend: even for the lowest flux levels (lowest values of
$\alpha$), pile-up is so severe that measured inner radii depart from
the input value by factors of a few.  The {\it Chandra}/ACIS
``continuous clocking'' mode is still overwhelmed at the baseline flux
level give in Table 2, and lines cannot be required in spectral fits
for $\alpha > 0.4$.

As with the ``low/hard'' state, the simulated {\it Chandra}/ACIS
``CC'' mode neutron star spectra were severely distorted by photon
pile-up, and other modes were less affected.  It was only possible to
detect and fit the iron line reliably for $\alpha = 0.1$ for {\it
Chandra}/ACIS ``continuous clocking'' mode (see Figure 5).  In the
case of the other simulated neutron star spectra, a limited number of
measured radii were 10\% smaller (e.g. $0.6~GM/c^{2}$) than the input
value.  This effect was only seen in simulations of spectra obtained
with the EPIC-pn camera in ``timing'' mode.  If these results are an
accurate characterization of the systematic errors incurred when
fitting mildly piled-up spectra with the EPIC-pn camera, it is a
modest systematic error.  Systematic uncertainties in the flux
calibration between different cameras on the same observatory are
approximately 7\% ({\it XMM-Newton} calibration document
XMM-SOC-CAL-TN-0083) and uncertainties between cameras on different
missions can easily exceed 10\% or more.

The neutron star spectra that were simulated assuming a narrower and
symmetric Gaussian emission line with E$=$6.7~keV and $\sigma =
0.33$~keV (as per Ng et al.\ 2010) also give clear results.  The
Gaussian line width, line centroid energy, and the normalization of
the disk blackbody component are plotted versus $\alpha$ in Figure 6.
(The disk blackbody component is the lowest-energy flux component in
the spectral model, and changes to its flux trace the degree of flux
redistribution due to photon pile-up.)  When photon pile-up is modest,
the line width and line energy are not strongly affected.  At the flux levels
simulated, the {\it XMM-Newton}/EPIC MOS ``full frame'' mode suffers
more severe pile-up.  The line becomes {\it narrower} as photon
pile-up (traced by the grade migration parameter $\alpha$) becomes
more severe (see Figure 6).  In no case do fits to the simulated
spectra measure a line that is falsely broad, asymmetric, nor
displaced to a significanly lower centroid energy.  In short, narrower
and symmetric lines are not observed to take on a relativistic shape
due to photon pile-up distortions.

The result that pile-up generally tends to produce {\it falsely
  narrow} lines can largely be understood in simple terms.  At high
  flux levels, photon pile-up will cause a CCD to register some low
  energy events as having a higher energy.  {\it Photon pile-up adds
  to the high energy continuum}.  Adding extra flux in the Fe K band
  has the effect of hiding the true profile of a relativistic line:
  the red wing blends with the continuum and the (relatively) narrow
  blue wing is all that remains.  The details of the input spectrum
  and the nature of a given CCD camera (its effective area curve, its
  effective frame time, how many event boxes tile an extraction
  region) are likely important.  This may partially explain why
  slightly different results are obtained from simulated EPIC-pn
  ``timing mode'' spectra of neutron stars.

The trends seen in Figure 4 are generally observed whenever pile-up is
important.  A broad range of incident spectra peak at approximately
1.0--1.5~keV, as do many detector effective area curves.  Pile-up of
photons to twice that energy, combined with a typical drop in
effective area at about 2~keV due abrupt changes in the mirror
reflectivity and also Si absorption in the CCD, causes a false excess
in the 2--3~keV band.  Depending on the specifics of the detector area
curve, this excess may be structured.  Pile-up also adds appreciably
through and above the Fe K band, causing the power-law to be falsely
hard and falsely high in flux.  This is seen as a steadily increasing
excess above 7~keV in the data/model ratio shown in Figure 4.  The
flux excess is greater when pile-up is more severe.  When fit with
typical spectral models, the flux excess at 2--3~keV and above 7~keV
act to create a false flux deficit just below the line energy, which
is again more severe as pile-up becomes more severe.  Owing to the
fact that more flux redistribution occurs when pile-up is more severe,
the full width of relativistic lines becomes increasingly difficult to
measure accurately.

\subsection{Pile-up and Disk Continua}

As shown in Figure 7, severe photon pile-up can have a strong impact
on efforts to study thermal emission from the accretion disk.
Depending on the specific detector and the input flux level (again,
traced by $\alpha$), pile-up can falsely reduce the flux in the disk
continuum by a factor of a few.  The radius inferred from fits with
the ``diskbb'' model depends on the square root of the flux
normalization.  More recent and physical disk models have more
parameters, but still measure the flux to determine an innermost
emission radius, and can be expected to be affected in the same
degree.  The results shown in Figure 7 depict the results of fits to
simulated ``high/soft'' state spectra, but consistent results are also
obtained in fits to the continuum in simulated spectra of other states
(see the bottom panel in Figure 6).

This effect may also be understood relatively simply.  Spectra of
bright X-ray binaries tend to peak at or near to the peak of the
effective area curve for X-ray telescopes with CCD spectrometers.
This is typically also the energy range in which the thermal disk
continuum dominates.  When pile-up is important, then, low energy flux
is preferentially lost from the accretion disk component due to grade
migration and flux redistribution.

It should be noted that even the baseline ``high/soft'' state flux
levels are actually quite high (see Table 1).  Simulations with
successively higher values of the grade migration parameter $\alpha$
trace successively higher flux levels.  Only the {\it Suzaku}/XIS with
a 1/4 window and 0.3~second burst option, and the {\it Swift}/XRT in
``windowed timing'' mode, are even marginally able to cope with this
flux level.  Even for these detectors and modes, strong mitigations
(e.g. annular extraction regions with large inner radii) would be
required to recover accurate spectral parameters.  The non-response of
the other detectors and modes to increases in $\alpha$ (see Figure 7)
merely indicates that the limits of the simulations are reached.  The
flux decrements indicated in Figure 7 for the {\it XMM-Newton}/EPIC MOS
in ``full frame'' mode, EPIC-pn in ``timing mode'', and the {\it
Chandra}/ACIS in ``continuous clocking'' mode, should not be taken as
realistic estimates of observed flux decrements.

\section{Discussion}
The results of the numerous simulations and fitting exercises detailed
above suggest that severe photon pile-up affects relativistic disk
spectra in clear and predictable ways:

Redistributing the low energy continuum to the high energy portion of
the spectrum causes relativistic emission lines to become {\it falsely
narrow}, giving a {\it falsely high} value for the inner disk radius.
In turn, this means that estimates of black holes spin parameters
based on the inner disk radius would be {\it falsely low}.  In spectra
where pile-up may be important and the success of mitigation is
uncertain, emission lines give {\it upper limits} on the radius of the
accretion disk, and {\it lower limits} on the value of the black hole
spin parameter.  The same trends are also observed in the case of
neutron stars, both when relativistic and simple narrow Gaussian line
functions are assumed.

Redistributing the low energy continuum to the high energy portion of
the spectrum has exactly the opposite impact on disk continua.  The
modified disk continuum gives {\it falsely low} values of the inner
disk radius, equating to {\it falsely high} values of black hole spin
based on that radius.  Therefore, when pile-up may be important,
and/or when efforts to mitigate pile-up may not have been entirely
successful, radii inferred from continuum fits are {\it lower limits}
and inferred spin parameters are {\it upper limits}.

When two diagnostics are skewed in the same sense, it is difficult to
detect a bias, and a potential check on derived quantities is lost.  The
results of the simple exercise undertaken in this paper are therefore
fortuitous: photon pile-up distorts relativistic disk lines and the
disk continuum in {\it opposing} ways.  Fitting both lines and the disk
continuum with relativistic spectral models and either checking that
the radii agree, or explicitly requiring agreement in the fit (see
Miller et al.\ 2009), will help to derive results that are robust
against pile-up distortions.  This procedure may be more difficult in
the case of neutron stars, however, because the disk continuum is
typically less distinct in neutron star spectra.

Efforts to understand the inner accretion flow geometry through
correlations between the flux in a disk line and the ionizing
continuum are thus also impacted by photon pile-up.  A particularly
interesting explanation of the flux trends seen in Seyfert-1 AGN such
as MCG-6-30-15 (Miniutti \& Fabian 2004) and NGC 4051 (Ponti et al.\
2006) is that gravitational light bending is altering the flux that
impinges on the disk from a power-law source of ionizing radiation.
In observations of such sources, photon pile up {\it is not} a
concern, but it might be a concern if CCD spectometers were to monitor
a stellar-mass accretor intensively.  It is worth noting that current
evidence of gravitational light bending in stellar-mass black holes is
drawn from gas spectrometer data that is unaffected by pile-up
(Miniutti, Fabian, \& Miller 2004; Rossi et al.\ 2005).

Two papers have recently commented on the possibility that photon
pile-up might create falsely broad emission lines in black hole
spectra.  In the first example, {\it Suzaku}/XIS spectra of GX 339$-$4
in an ``intermediate'' state were extracted from different annuli and
compared (Yamada et al. 2009).  Spectra from larger annuli (presumably
suffering from less pile-up) are not found to strongly require black
hole spin.  Yamada et al.\ (2009) suggest that pile-up may influence
the line shape, but fitting results are consistent with a very broad
line and a spinning black hole at the 90\% level of confidence (see
Miller et al.\ 2008).  The relatively small number of counts in the
wings of the PSF can partially account for the lack of a strong spin
requirement and lower statistical certainty.  A separate but equally
important issue is that the disk reflection model used by Yamada et
al.\ (2009) was not convolved with the line element expected for
emission from the inner disk; this is physically inconsistent in that
it implies a stationary, non-orbiting reflector.  The immediate effect
is to falsely add to the continuum in the vicinity of the emission
line, falsely narrowing the line.

Recent work by Done \& Diaz Trigo (2010) examined {\it XMM-Newton}
spectra of GX 339$-$4 in a ``low/hard'' spectral state.  Spectra from
the MOS cameras were found to suffer from pile-up, even when
extracting counts in annuli, whereas the EPIC-pn ``timing'' mode
spectra were claimed to be free of distortions from photon pile-up.
The pn spectra were also found to deliver narrower line profiles than
the MOS spectra, implying a larger inner disk radius compared to
values reported in prior work (e.g. Miller et al.\ 2004; Reis et al.\
2008; also see Wilkinson \& Uttley 2009).  Repeating the MOS
extraction exactly as detailed in Reis et al.\ (2008), it is apparent
that the disk line profile does not vary with the inner radius of
annular extraction regions (see Figure 8).  The line profile in the
annulus identified by Done \& Diaz Trigo (2009) as being largely free
of pile-up, closely matches the line profiles seen in spectra
extracted from smaller annuli, when each spectrum is allowed to have
its own continuum.  This means that the radii measured in the MOS
spectra are not distorted by photon pile-up, but that the radii
derived in fits to the pn spectra are distorted.

This conclusion is echoed by our simulations.  The results detailed
above strongly suggest that the fact of a narrower line profile in the
pn means that it is actually piled-up and does not measure the true
line width.  Indeed, the departure of 2--3~${\rm GM}/{\rm c}^{2}$
shown in Figure 2 is commensurate with the radius difference between
fits to the MOS spectra reported by Miller et al.\ (2006) and Reis et
al.\ (2008), and the pn spectrum as fit by Wilkinson \& Uttley (2009).
The data/model ratio in Figure 7 of Done \& Diaz Trigo (2009) shows a
flux excess at 2~keV and a flux decrement between 3 keV and the Fe K
line, very similar to the trends shown in Figure 4 in this paper.  The
observed spectra and our simulation results both show that photon
pile-up makes lines artificially narrow.

Following Done \& Diaz Trigo (2010), Ng et al.\ (2010) have analyzed a
number of spectra of neutron star low-mass X-ray binaries, observed
with the {\it XMM-Newton}/EPIC-pn camera in ``timing'' mode.  Their
analysis suggests that the spectra suffer from a degree of photon
pile-up.  When the center of the PSF is excluded, the line profiles
are found to be more consistent with Gaussian profiles.  Ng et al.\
(2010) conclude that pile-up acted to falsely create skewed,
relativistic line profiles.  This conclusion runs counter to the
results of our simulations, which show that (1) severe pile-up acts to
falsely narrow both relativistic and simple Gaussian lines, not to
falsely broaden lines; and (2) pile-up distortions to typical neutron
star line spectra are not expected to be extreme (see Figure 5) for
sources with flux levels similar to Serpens X-1, except for {\it
Chandra}/ACIS ``continuous clocking'' mode.

Independent {\it Suzaku} and {\it XMM-Newton} observations of Serpens
X-1 reveal relativistic Fe lines with a clear red wing (see Figure 9;
also see Bhattacharyya \& Strohmayer 2007, Cackett et al.\ 2008, Cackett et
al.\ 2010).  It is already clear that {\it Suzaku} line profiles in
this source and similar sources do not depend on the inner radius of
event extraction annuli (Cackett et al.\ 2010; see Figure 11) -- the
relativistic profiles do not result from photon pile-up.  The nature
of the line seen with {\it Suzaku} strongly suggests that the red wing
seen in the {\it XMM-Newton} spectrum is due to dynamical
broadening and red-shifting, not photon pile-up.

We re-reduced the {\it XMM-Newton} pn observation of Serpens X-1
exactly as detailed in Bhattacharyya \& Strohmayer (2007), and
extracted spectra including and excluding the center of the PSF.  The
primary effect of excluding the center of the PSF is merely to reduce
the number of photons in the resultant spectrum, thereby lowering the
sensitivity to the point that the red wing cannot be detected.  The
line is neither narrower nor weaker when the center of the PSF is
excluded -- it is simply less defined (see Figure 10).  Ng et al.\
(2010) do not present direct comparisons of line profiles in spectra
obtained in dfferent extraction regions.

Given that: (1) {\it Suzaku} spectra of neutron stars reveal
relativistic lines that are clearly not due to photon pile-up
distortions, (2) a number of {\it Suzaku} and {\it XMM-Newton} line
profiles are remarkably similar (e.g. SAX J1808.6$-$3808, see Figure 3
in Cackett et al.\ 2010; GX 349$+$2, see Figure 9 in Cackett et al.\
2010), (3) the effect of excluding the center of the PSF in the {\it
XMM-Newton} spectrum of Serpens X-1 is merely to reduce definition in
the Fe line (not its width nor its strength), and (4) our simulations
show that severe pile-up acts to artificially {\it narrow} both
relativistic and simple Gaussian lines, it is likely that dramatic
reductions in sensitivity drove the results obtained by Ng et al.\
(2010) and led to faulty conclusions regarding the ability of pile-up
to create false relativistic line profiles.

The fact that sensitive spectra are required to detect the red wings
of relativistic line profiles is not a revelation.  This fact fueled
the requirement of minimum sensitivity thresholds in recent surveys of
relativistic lines in Seyfert AGN (see Nandra et al.\ 2007).  Recent
work on neutron star spectra makes the necessity of sensitive spectra
even more clear.  Deep observations of a source such as GX 349$+$2
with {\it Suzaku} reveal a relativistic line profile, whereas short
observations with the {\it Chandra}/HETGS -- which has a lower
collecting area -- only recovers the narrower blue wing of the line
profile (see Figure 3 in Cackett et al.\ 2009b).  The {\it Chandra}
spectrum is fully consistent with the relativistic line profile found
using {\it Suzaku}, it simply does not have the sensitivity needed to
detect the red wing against the continuum.

The above discussion is mostly focused on {\it XMM-Newton} and {\it
Suzaku} spectroscopy, because the large effective area of these
missions ensures that they are well-suited to relativistic
spectroscopy.  {\it Swift} is highly flexible, but its smaller
collecting area and short observations mean that typical spectra lack
the sensitivity to detect and measure relativistic disk lines well.
However, {\it Swift} is very well-suited to black hole spin
measurements using the disk continuum, and our results suggest that
even in ``windowed timing'' mode, annular extraction regions are
needed to avoid spectral distortions due to photon pile-up (see Figure
6).  Brief discussions of pile-up and excluding the central portion of
the PSF in ``windowed timing'' mode are given in Rykoff et al.\ (2007)
and Rykoff, Cackett, \& Miller (2010).

Two combinations of detectors and modes are not treated in this work
{\it because} they represent effective means of preventing photon
pile-up.  As noted previously, the ``burst'' mode of the EPIC-pn
camera aboard {\it XMM-Newton} may provide the {\it best} means of
preventing photon pile-up, as its short read-out time is suited to
extremely bright sources.  This comes at a cost in sensitivity,
however, as ``burst'' mode has a livetime fraction of just 0.03.
Pile-up can also be avoided by observing with the {\it Chandra}/HETGS,
which disperses a spectrum onto the ACIS spectrometer.  Pile-up can be
further avoided by reading-out the dispersed spectrum in ``continuous
clocking'' mode.  

{\it Chandra}/HETGS observations of the stellar-mass black hole
GX~339$-$4 in an ``intermediate'' state measured an inner disk radius
of $1.3_{-0.1}^{+1.7}~{\rm GM}/{\rm c}^{2}$ (Miller et al.\ 2004b).
This analysis used the same relativistically-blurred reflection model
applied to a later observation in the same state using {\it Suzaku}
(Miller et al.\ 2008).  Analysis of the later observation found a spin
parameter of $a = 0.89\pm 0.04$, correspinding to ${\rm r} = 2.0\pm
0.2~{\rm GM}/{\rm c}^{2}$ (Bardeen, Press, \& Tuekolsky 1972).  The
radius values measured in GX~339$-$4 are consistent in the two
separate observations, using very different detectors.  Similarly,
{\it Chandra}/HETGS observations of the neutron star X-ray binary 4U
1705$-$44 find a very broad line, with a Gaussian width (FWHM) of $1.2\pm
0.2$~keV; when fit with a relativistic diskline model, a radius of
${\rm r} = 7^{+4}_{-1}~{\rm GM}/{\rm c}^{2}$ is measured (Di Salvo et
al.\ 2005).  Here again, the line properties found using grating
spectra -- which avoid severe pile-up -- are consistent with those
found using {\it Suzaku} and {\it XMM-Newton} (Reis, Fabian, \& Young
2009, Di Salvo et al.\ 2009, Cackett et al.\ 2010), and inconsistent
with the much narrower lines found by Ng et al.\ (2001) when
extracting only the wings of {\it XMM-Newton}/EPIC-pn ``timing'' mode
spectra.

An {\it XMM-Newton}/EPIC-pn ``burst'' mode observation of
GRS~1915$+$105 in a ``plateau'' state (similar to the low/hard state
in most black holes) may also validate the results of our simulations
and the discussion above.  Martocchia et al.\ (2006) observed GRS
1915$+$105 twice in the ``plateau'' state -- once in ``timing'' mode
and once in ``burst mode'' -- at consistent flux levels.  The
``timing'' mode spectrum may suffer from modest distortions due to
photon pile-up.  The power-law index was measured to be harder than in
the ``burst'' mode observation ($\Gamma = 1.686^{+0.008}_{-0.012}$
versus $\Gamma = 2.04^{+0.01}_{-0.02}$), indicative of possible
pile-up.  The iron line is found to be fairly narrow in the ``timing''
mode observation: the line is only visible above $\sim$6.2~keV when
the spectrum is fit with a power-law, the inner radius is constrained
to be greater than 240~${\rm GM}/{\rm c}^{2}$, and a reflection
fraction of ${\rm R} = 0.35^{+0.02}_{-0.02}$ is measured.  In
contrast, when observed using ``burst'' mode, the line is visible down
to 5~keV (or lower) relative to a simple continuum, a radius of less
than $20~{\rm GM}/{\rm c}^{2}$ is required, and a reflection fraction
of $R=1.69^{+0.16}_{-0.04}$ is measured (Martocchia et al.\ 2006).
(This {\it XMM-Newton}/EPIC-pn ``burst'' mode spectrum is broadly
conistent with a {\it Suzaku} observation of GRS 1915$+$105 in the
``plateau'' state.  In that spectrum, the line is also broad, and
consistent with the ISCO; see Blum et al.\ 2009.)  These results
square with a central result of the simulations presented in this
paper: photon pile-up acts to make relativistic lines appear to be
falsely narrow.

Though an investigation is beyond the scope of this paper, it should
be noted that efforts to mitigate pile-up can also distort spectra.
The PSF of an X-ray telescope is energy-dependent.  Extracting events
in annuli may avoid the piled-up core of a given PSF, but the spectrum
is then derived from a portion of the PSF that is not calibrated as
well as the core.  Moreover, depending on the PSF, annular regions may
only extract a vanishing fraction of the incident photon flux,
complicating the detection of weak spectral lines.  Relativistic disk
lines are often 10--20\% features above the continuum; extracting
spectra from regions of the PSF from which the energy and flux
calibration are not known to much better than 10\% could easily make
it difficult to recover the details of a given line profile.
Depending on how much flux is excluded in the annulus, limited photon
statistics will also serve to complicate the detection of line asymmetry.

Finally, it is worth emphasizing that our results depend on a specific
combination of mirror technology and detector technology.  Presently,
CCD spectrometers sit at the focus of gold foil or gold-coated mirors;
this partially accounts for an overall telescope efficiency curve that
is highest below 2 keV.  New mirror technology, such as Si pore optics
(e.g. Beijersbergen, M., et al., 2004), may produce different CCD
photon pile-up effects, and distortions to disk continua and disk
lines may not longer skew in the opposite sense.

\section{Conclusions}

Extensive simulations of photon pile-up for a number of spectral forms
and detector parameters suggest that severe pile-up can distort
spectroscopic signatures of the inner accretion disk, in largely
predictable ways.  The results of our work can be summarized as follows:\\

\noindent{$\bullet$ The degree of photon pile-up in a spectrum depends
  on the input flux level from a source and the effective area of the
  telesecope, but it is also depends on the number of event boxes
  tiling the PSF and the frame time of the CCD.}

\noindent{$\bullet$ The relevant time scale for pile-up considerations
  is the time required to clock a full event box, not merely one row
  of charge.}

\noindent{$\bullet$ Tiling the PSF with many event boxes and short
  frame times are two independent means of reducing photon pile-up.
  However, when short frame times are achieved by changing the size of
  an event box and under-sampling the PSF, pile-up mitigation is
  partially compromised.}

\noindent{$\bullet$ Flux redistribution due to photon pile-up causes
  emission lines -- whether relativistic or symmetric and
  intrinsically narrow -- to be observed as falsely narrow.  Measured
  inner disk radii are then falsely large, and inferred black hole
  spin parameters are falsely low.  Relevent observed spectra, though
  small in number, support this finding.}

\noindent{$\bullet$ Grade migration due to photon pile-up causes the
  low energy spectrum -- principally the disk component -- to have a
  falsely low flux.  Measured inner disk radii are therefore falsely
  small and inferred spin parameters are falsely high.}

\noindent{$\bullet$ Deriving inner disk radii and/or black hole spin
  parameters by linking that parameter to a joint value in both disk
  continuum and disk reflection models may yield more robust results
  when photon pile-up cannot be avoided or successfully mitigated.}

%\section{Acknowledgements}

\hspace{0.1in}

We wish to acknowledge the referee, Keith Jahoda, for a thoughtful
review of this work and for comments that improved the paper.  We
thank John Davis, Matthias Ehle, Lothar Str{\"u}der, J{\"o}rn Wilms,
and Maria Diaz-Trigo, Mark Reynolds, Dipankar Maitra, and Tiziana Di
Salvo, and Giorgio Matt for helpful discussions.

\pagebreak

% \centerline{~\psfig{file=f1.ps,width=3.2in}~}
% \figcaption[h]{\footnotesize A power density spectrum of GX~339$-$4
% obtained with {\it RXTE} during the long {\it XMM-Newton} observation
% is shown above.  This power spectrum is typical of the low--hard
% state, in that it shows high fractional variability and
% band-limited noise.  The power spectrum was fit with three Lorentzians
% ($\chi^{2}/\nu = 204/177$) shown with dashed and dotted lines.}
% \medskip

\begin{table}[t]
\caption{Input Model Parameters}
\begin{footnotesize}
\begin{center}
\begin{tabular}{lllll}
Parameter & Very High / Intm. & Low/Hard &  High/Soft & Z/Atoll	\\
\tableline
${\rm N}_{\rm H}~(10^{22}~{\rm cm}^{-2})$ &  0.5	& 0.5 & 0.5 & 0.5 \\

${\rm kT}_{\rm disk}$ (keV) & 0.76 & 0.18 & 0.97 & 1.21 \\	

${\rm K}_{\rm disk}$ & 2300.0 & 160000.0 & 7050.0 & 103.0 \\

${\rm kT}_{\rm bbody}$ (keV) & -- & -- & -- & 2.28 \\

${\rm K}_{\rm bbody}$ ($10^{-2}$) & -- & -- & -- & 4.9 \\

$\Gamma$ & 2.60	& 1.60	& 2.48	& 3.60	\\

${\rm K}_{\rm pl}$ & 2.20 & 0.50	& 3.38	& 0.18	\\

${\rm E}_{Laor}$ (keV) & 6.7 & 6.7 & --	& 6.7	\\

$q$ & 3.0 & 3.0	& -- & 3.0 \\

${\rm r}_{in} (\rm GM/c^{2})$ & 6.0 & 6.0 & -- & 6.0 \\

$i$ (deg.) & 30.0 & 30.0 & -- &	30.0 \\

${\rm K}_{\rm line} (10^{-3})$ & 7.7 & 8.5 & -- & 8.0 \\

\tableline

Flux ($10^{-9}~{\rm erg}~{\rm cm}^{-2}~{\rm s}^{-1}$) & 36.9 & 2.9 & 86.2 & 6.0 \\

\tableline
\end{tabular}
\vspace*{\baselineskip}~\\ \end{center} 
\tablecomments{The table above lists the input spectral parameters
  used to generate the simulations considered in this work.  The
  spectral parameters were chosen to be representative of different
  spectral states in stellar-mass Galactic black holes, and a typical
  medium-intensity state in an ``atoll'' neutron star X-ray binary.
  The ``very high'' or ``intermediate'' state paramters are based on
  an observation of GX 339$-$4 as reported by Miller et al.\ (2004);
  the ``low/hard'' state parameters are based on observations of GX
  339$-$4 as reported by Miller et al.\ (2006), Tomsick et al.\
  (2008), and Wilkinson \& Uttley (2009); the ``high/soft'' state
  parameters are taken from an observation of 4U 1543$-$475 made on
  MJD 52452 as reported by Park et al.\ (2004), and the typical
  Z/Atoll spectrum is based on an observation of Serpens X-1 from
  Cackett et al.\ (2008).  In the case of the neutron star spectrum,
  three continuum components (a disk blackbody, a simple blackbody,
  and a power-law) are needed to model the continuum.  The observed
  (absorbed) flux is given for each input spectrum.}
\vspace{-1.0\baselineskip}
\end{footnotesize}
\end{table}

\clearpage

\begin{table}[t]
\caption{Detectors, Modes, and Parameters}
\begin{footnotesize}
\begin{center}
\begin{tabular}{llllllllll}
\tableline
\multicolumn{2}{l}{Mission/ Instrument \& Mode} &  ${\rm T}_{\rm R}^{a}$  & Radius$^{b}$ & Pixel Size$^{c}$ & Event Box$^{d}$ & ${\rm T}_{\rm eff.}^{g}$ & Regions$^{h}$ \\
\multicolumn{2}{l}{~} & ($10^{-3}$~s)  & (arcsec) & (arcsec) & (pixels) & ($10^{-3}$~s) & ~ \\
\tableline
\multicolumn{2}{l}{Chanda/ACIS ``continuous clocking''} & 2.85 & 2.5 & 0.5 & 3x3 & 8.55 & 3.3 \\

\multicolumn{2}{l}{Swift/XRT ``windowed timing''} & 1.77 & 9.0 & 2.36 & 10x7 & 1.77 & 1.1 \\

\multicolumn{2}{l}{XMM-Newton/pn ``timing''} & 0.03 & 120 & 4.1 & 30x3 & 0.09 & 1.0 \\

%\multicolumn{2}{l}{XMM-Newton/pn ``full frame''} & 73.4 & 20 & 4.1 & 3x3 & 220 & 8.3 \\

\multicolumn{2}{l}{XMM-Newton/MOS ``full frame''} & 2600 & 20 & 1.1 & 3x3 & 2600 & 115.3 \\

\multicolumn{2}{l}{Suzaku/XIS ``1/4 Window$+$0.3s Burst''} & 300 & 60.0 & 1.0 & 3x3 & 300.0 & 1256.0 \\

\tableline
\end{tabular}
\vspace*{\baselineskip}~\\ \end{center} 
\tablecomments{The table above lists important features of the
detectors and modes considered in our simulations, as well as the
details of the extraction regions that were assumed.  Please see
Section 6 for a detailed discussion.\\
$^{a}$ ${\rm T}_{\rm R}$ is the time resolution of a given instrument
and mode.  In standard imaging modes, this is the ``frame time'' over
which the CCD records data.  In specialized timing modes, ${\rm T}_{\rm
R}$ is actually the time to transfer one row (of pixels or
macropixels) of charge.  \\
$^{b}$ This column gives the ``radius'' of the extraction regions used
  in simulating an observation using a given detector and mode.  For
  Chandra, the PSF is sufficiently small that the 90\% encircled
  energy radius was used.  For other missions, nominal extraction
  regions were selected to reflect the radius over which the encircled
  energy changes approximately linearly with radius.  In the case of
  {\it Suzaku} and {\it Swift}, the radii used are essentially the
  half-power radii.  The radii used for the {\it XMM-Newton} MOS and
  pn cameras encircle approximately 70\% and 75\% of the total
  incident energy, respectively.\\
$^{c}$ This column gives the nominal pixel size for each detector.\\
$^{d}$ This column gives the size of an event box for each detector
and mode considered.  The {\it Suzaku} mode and {\it
XMM-Newton}/EPIC-MOS modes considered in this work are imaging modes
wherein standard pixels and event boxes are retained.  The {\it
Chandra}/ACIS ``continuous clocking'' mode is a specialized timing
mode but retains the standard 3x3 event box used in imaging modes.
The {\it XMM-Newton}/EPIC-pn ``timing'' and {\it Swift}/XRT ``windowed
timing'' modes achieve a fast read-out partially be creating
``macropixels'' that are 30 pixels and 10 pixels, respectively, in the
read-out direction.  The {\it Swift} ``windowed timing'' mode is
unique in that the event box is changed from a 3x3 format to a 10x7
format.\\
$^{g}$ ${\rm T}_{\rm eff.}$ is the effective time resolution for a
given instrument and mode.  In specialized timing modes, this is the
time to clock one full event box of charge, not merely one row of
charge.\\
$^{h}$ This field notes the number of event boxes that cover the
  extraction radius.}
\vspace{-1.0\baselineskip}
\end{footnotesize}
\end{table}

\clearpage

\centerline{~\psfig{file=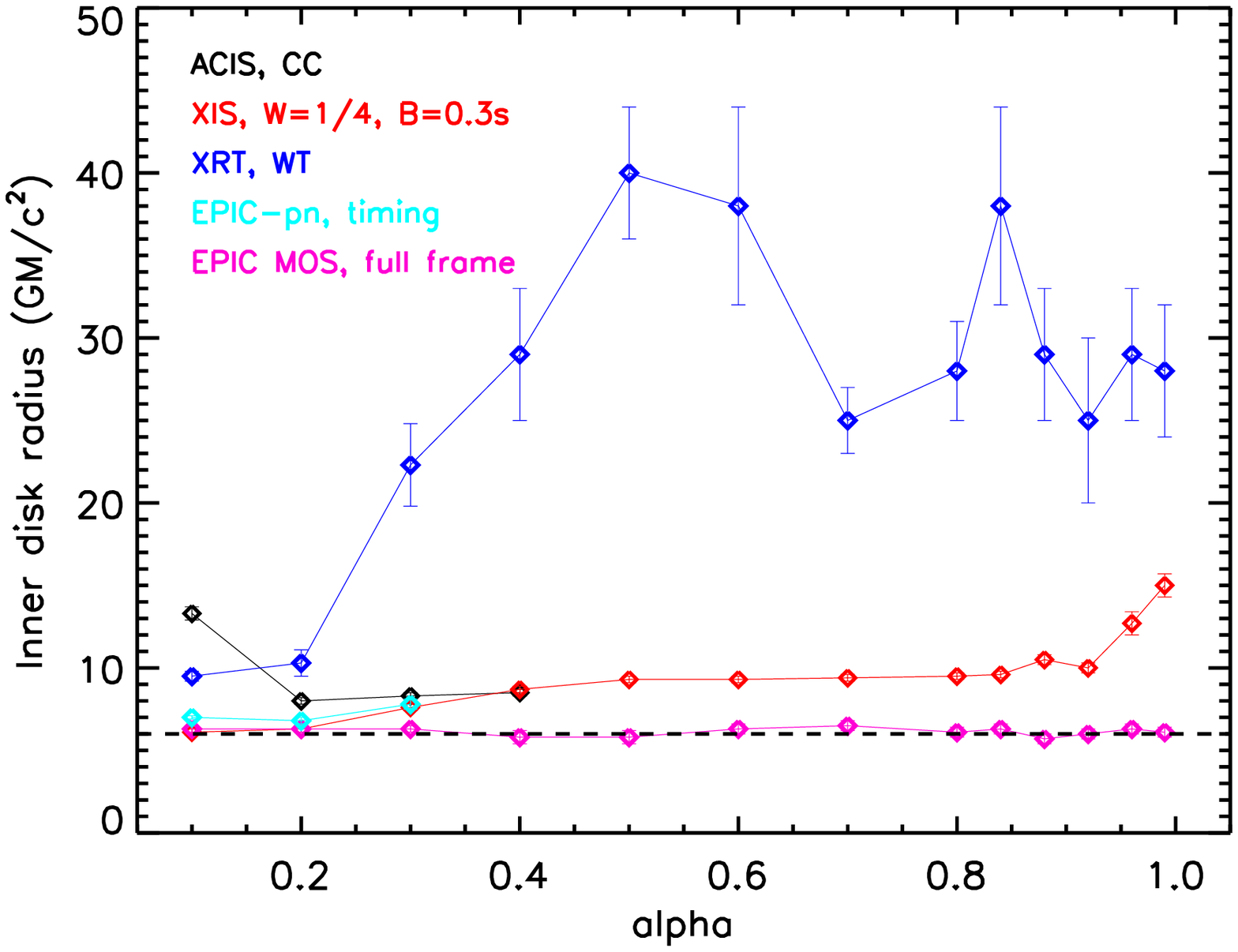,width=5.0in}~}
\figcaption[h]{\footnotesize The plot above depicts the evolution of
  the inner disk radius measured via a relativistic iron emission line
  (in units of ${\rm GM}/{\rm c}^{2}$) versus the severity of photon
  pile-up (governed by the grade migration parameter $\alpha$), for
  simulated spectra assuming a form typical of the ``very high'' state
  in accreting black holes (see Section 5 and Table 1).  The dashed
  horizontal line at $6~{\rm GM}/{\rm c}^{2}$ denotes the input radius
  used in all simulated spectra.  An iron line could not be required
  in fits to simulated {\it Chandra}/ACIS ``continuous clocking'' mode
  spectra past $\alpha = 0.92$, nor above $\alpha > 0.3$ for fits to
  simulated EPIC-pn spectra.  Fits to simulated EPIC MOS spectra do
  not show an evolution in radius because pile-up is so severe that
  event loss dominates over grade migration distortions; this is a
  limitation of our simulations and it does not suggest that the MOS
  should be operated in ``full frame'' to observe a source this
  bright.  The results above suggest that the {\it Suzaku}/XIS (with a
  1/4 window and 0.3 seconds burst option) and {\it Swift}/XRT (in
  windowed timing mode) are capable of dealing with flux levels this
  high but would give falsely large disk radii unless pile-up
  distortions were mitigated (e.g. by extracting counts in annuli).}
\medskip

\clearpage

\centerline{~\psfig{file=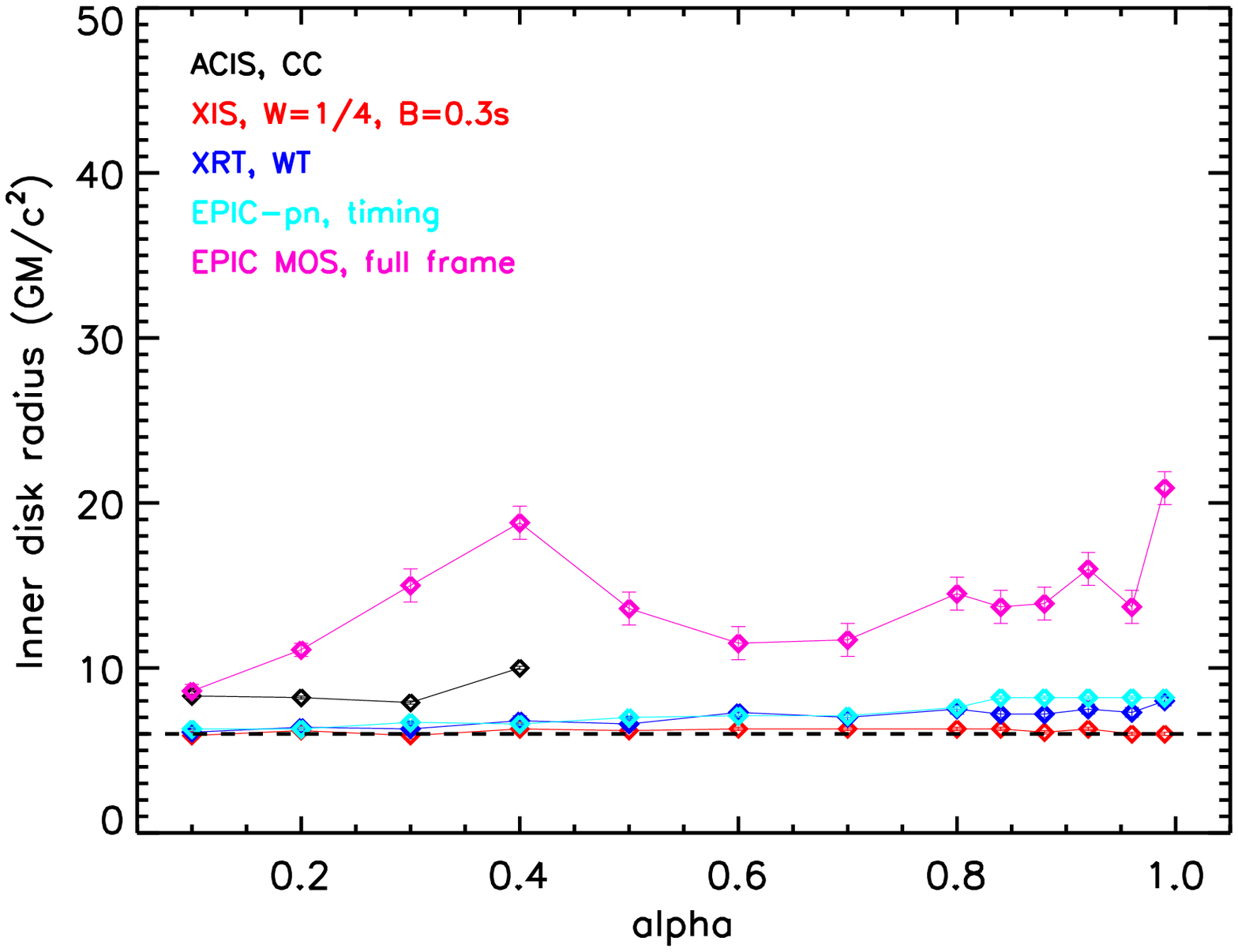,width=5.0in}~}
\figcaption[h]{\footnotesize The plot above depicts the evolution of
  the inner disk radius measured via a relativistic iron emission line
  (in units of ${\rm GM}/{\rm c}^{2}$) versus the severity of photon
  pile-up (governed by the grade migration parameter $\alpha$), for
  simulated spectra assuming a form typical of the ``low/hard'' state
  in accreting black holes (see Section 5 and Table 1).  The dashed
  horizontal line marks the input radius of $6~{\rm GM}/{\rm c}^{2}$.
  An iron line could not be required in fits to simulated {\it
  Chandra}/ACIS ``continuous clocking'' mode spectra past $\alpha =
  0.4$.  The results shown above suggest that the {\it Suzaku}/XIS
  (with a 1/4 window and 0.3 seconds burst option) can potentially
  deliver a nearly nominal response and accurate radii at flux levels
  typical of the ``low/hard'' state.  In contrast, our results suggest
  that radii derived from spectra obtained using other detectors and
  modes could be to high by approximately 30\%, unless additional
  mitigations were adopted.}
\medskip

\clearpage

\centerline{~\psfig{file=f3.ps,width=6.0in,angle=-90}~}
\figcaption[h]{\footnotesize The plot above illustrates the effect of
  pile-up on the spectral continuum.  The spectra shown above were fit
  using the parameters of the input spectral model.  In this case,
  simulated ``very high'' state inputs were used, and
  observations with the {\it Suzaku}/XIS responses assuming a 1/4
  window option and 0.3~second burst option are shown.  Values of the
  grade migration parameter $\alpha = 0.1, 0.4, 0.7, 0.96$ are shown
  above in black, red, green, and blue, respectively.  As pile-up
  becomes more severe, the high energy continuum becomes harder due to
  event grade migration.}
\medskip

\clearpage

\centerline{~\psfig{file=f4.ps,width=6.0in,angle=-90}~}
\figcaption[h]{\footnotesize The plot above illustrates the effect of
  pile-up on iron lines.  As pile-up becomes more severe, an excess
  grows between 2-3 keV, a flux decrement can be seen between 3~keV
  and 5~keV, and the line becomes narrower.  This is due to a
  combination of flux redistribution, grade migration, and event loss
  altering the shape of the continuum spectrum.  The plot above was
  made by fitting simple disk plus power-law continua to simulated
  ``very high / intermediate'' state spectra using the Suzaku XIS
  responses and assuming a 1/4 window and 0.3 second burst option.
  The 4--7~keV region was ignored when fitting the continuum, and then
  included when making the data/model ratio plot shown above.  Values
  of the grade migration parameter $\alpha = 0.1, 0.4, 0.7, 0.96$ are
  shown above in black, red, green, and blue, respectively.  Errors
  have been neglected for visual clarity.}
\medskip

\clearpage

\centerline{~\psfig{file=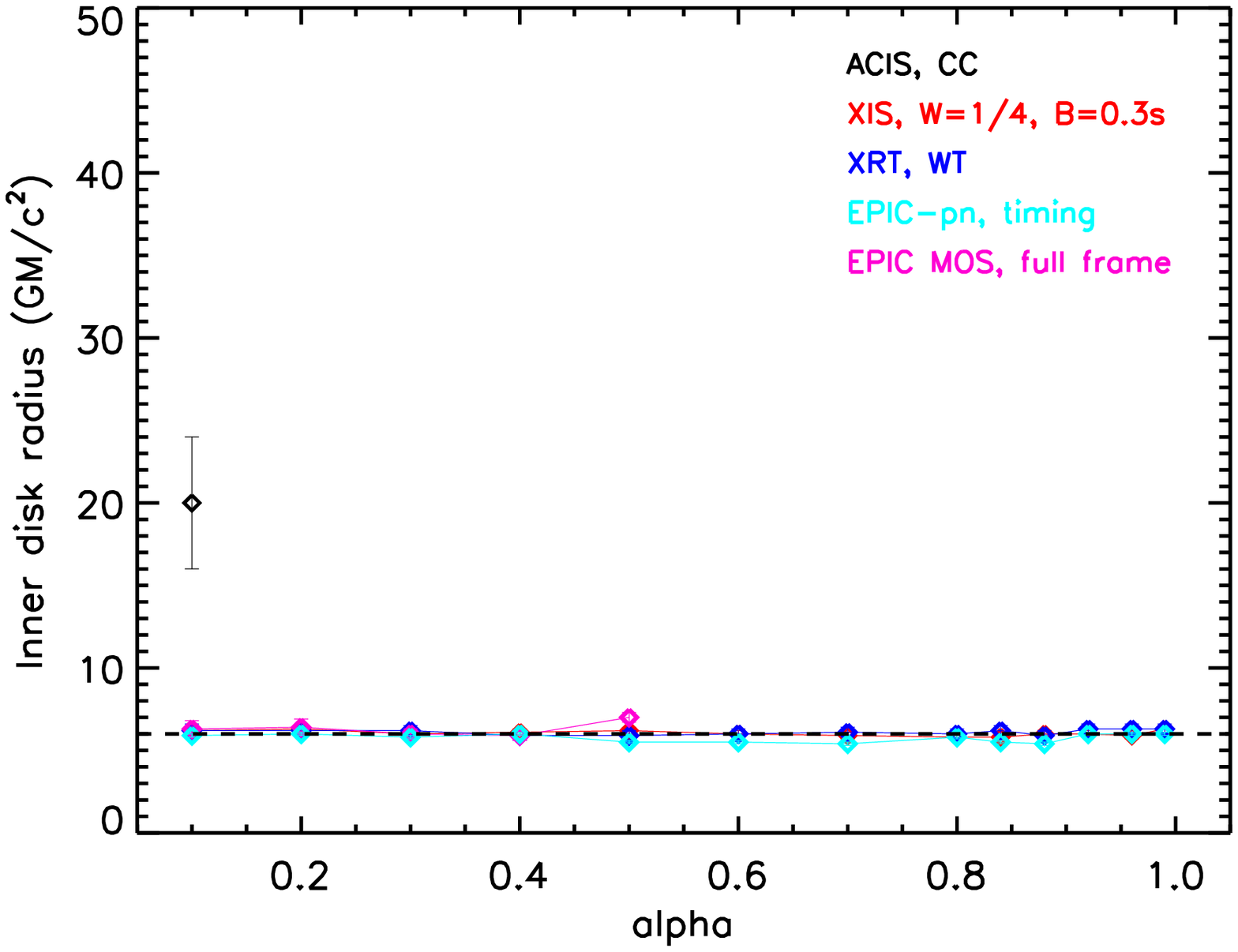,width=5.0in}~}
\figcaption[h]{\footnotesize The plot above depicts the evolution of
  the inner disk radius measured via a relativistic iron emission line
  (in units of ${\rm GM}/{\rm c}^{2}$) versus the severity of photon
  pile-up (governed by the grade migration parameter $\alpha$), for
  simulated spectra assuming a form typical of ``Z'' and ``atoll''
  neutron star X-ray binaries (see Section 5 and Table 1).  The dashed
  horizontal line marks the input radius of $6~{\rm GM}/{\rm c}^{2}$.
  An iron line could not be required in fits to simulated {\it
  Chandra}/ACIS ``continuous clocking'' mode spectra past $\alpha =
  0.1$.  The results shown above suggest that other detectors and
  modes can deliver a nearly nominal response and accurate inner disk
  radii at the flux level assumed in our simulations.}
\medskip

\clearpage

\centerline{~\psfig{file=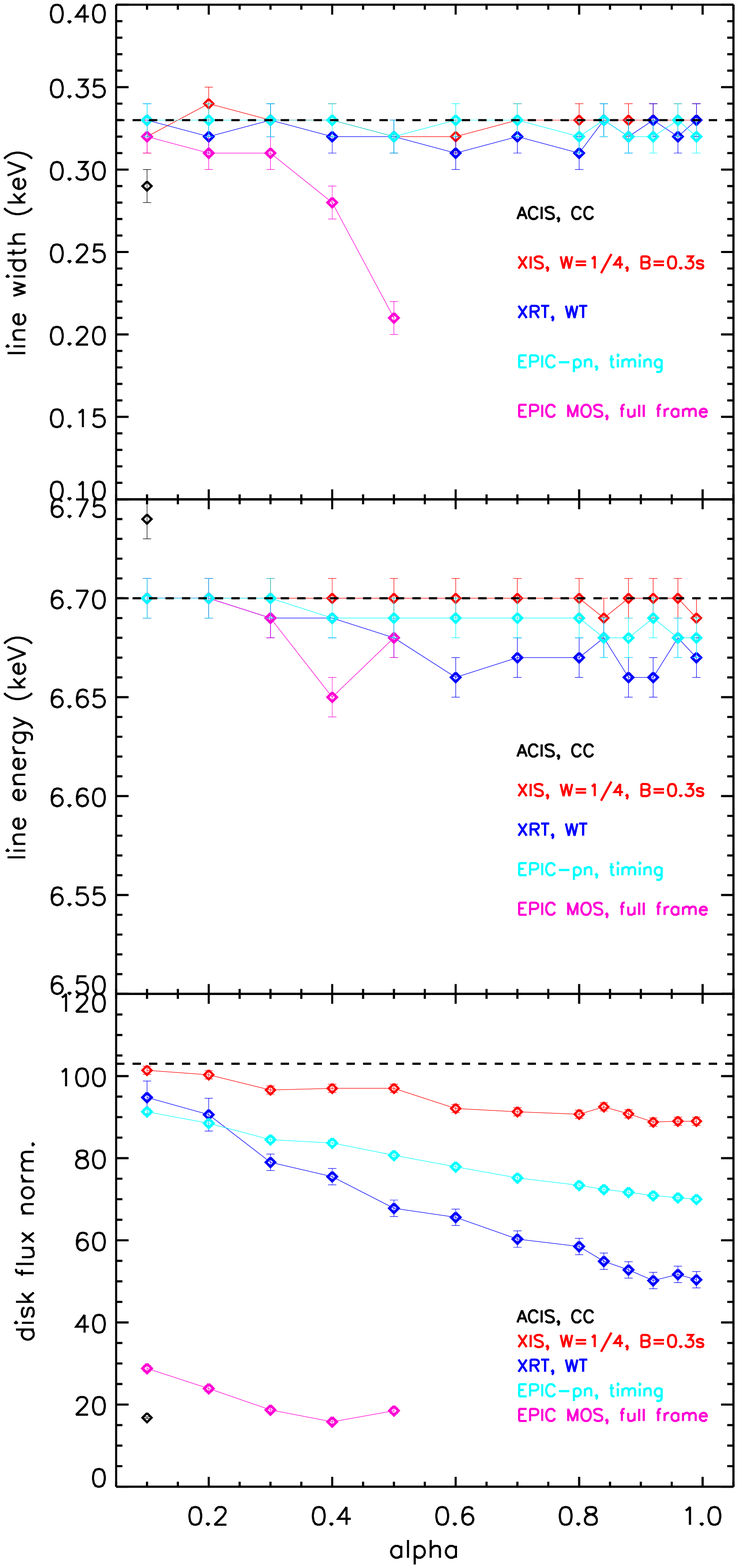,width=4.0in}~}
\vspace{-0.5in}
\figcaption[h]{\footnotesize The plot above shows the results of
replacing a relativistic line profile with a narrow Gaussian line
(E$=$6.7~keV, $\sigma=0.33$~keV) in simulated ``Z'' and ``atoll''
neutron star X-ray binary spectra.  The line width (top), line energy
(middle), and disk blackbody flux normalization (bottom) are plotted
as a function of pile-up severity, as traced by the grade migration
parameter $\alpha$.  Nominal input (simulated) values for each
parameter are indicated by dashed horizontal lines (see Table 1).  The
disk blackbody is the lowest enegy component in the spectral model,
and the evolution of its flux normalization traces the degree of flux
redistribution due to photon pile-up.  The plots above clearly show
that intrinsically narrow lines are not artificially broadened nor
shifted to significantly lower energy by photon pile-up distortions.
When pile-up is severe, intrinsically narrow lines are measured to be
even narrower.}
\medskip

\clearpage

\centerline{~\psfig{file=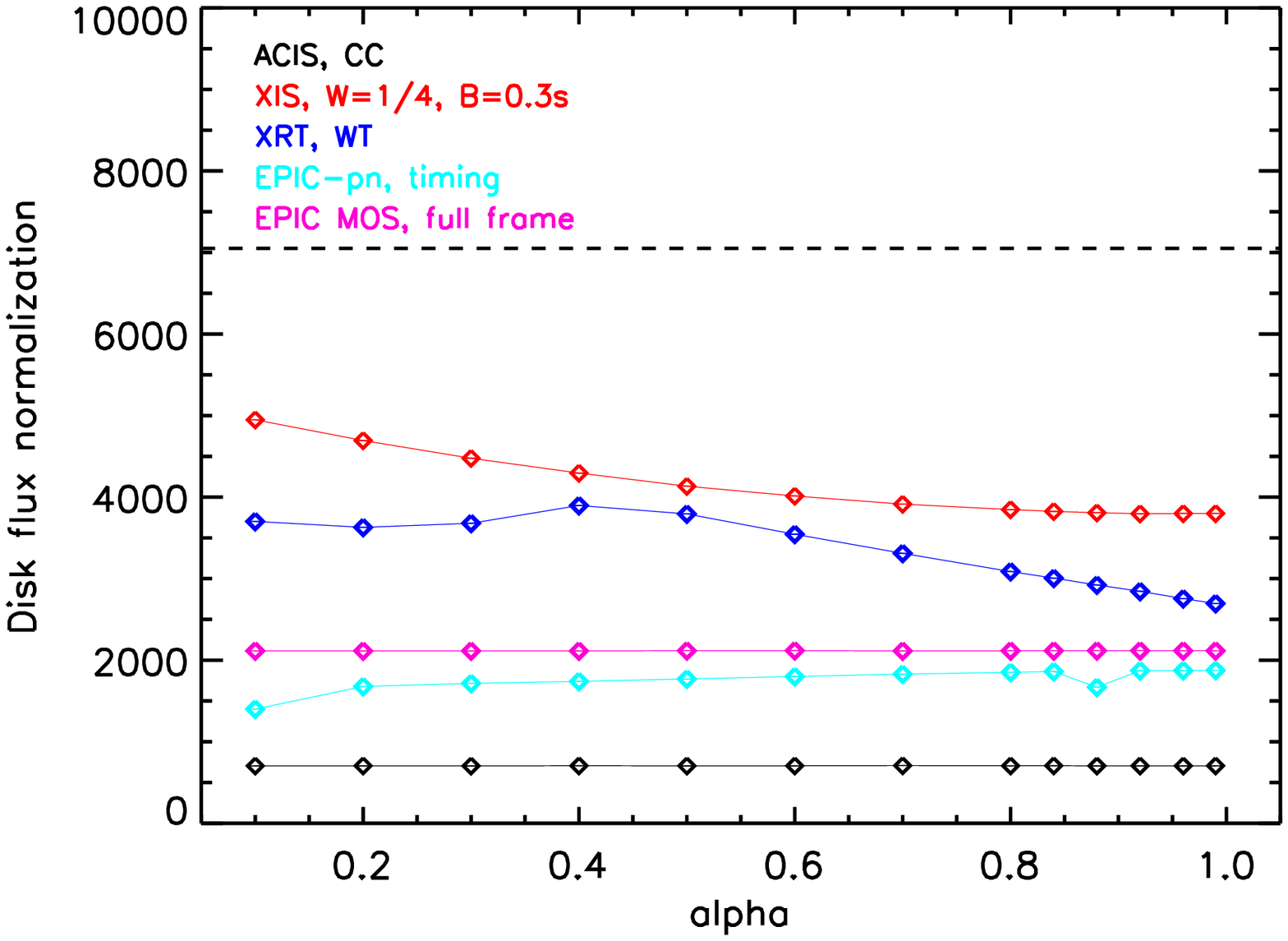,width=5.0in}~}
\figcaption[h]{\footnotesize The plot above depicts the evolution of
  the disk continuum flux normalization versus the severity of photon
  pile-up (indicated by the grade migration parameter $\alpha$), for
  simulated spectra assuming a form typical of the ``high/soft'' state
  in accreting black holes (see Section 5 and Table 1).  The dashed
  horizontal line marks the input flux normalization level.  The fact
  that the normalization is independent of $\alpha$ in the {\it
  XMM-Newton}/EPIC pn and MOS spectra and {\it Chandra}/ACIS spectra
  is an artifact of the simulations; it indicates that these modes are
  overwhelmed by the high flux typical of the ``high/soft'' state and
  none of these three modes should be regarded as better suited than
  the others in this regime.  The flux levels indicated for these
  detectors and modes should not be regarded as realistic estimates of
  the flux decrement that would be observed.  In contrast, the {\it
  Suzaku}/XIS (with a 1/4 window and 0.3~seconds burst option) and the
  {\it Swift}/XRT (in windowed timing mode) may be suited to this
  state but will give falsely small disk flux normalizations unless
  additional mitigations are taken.  For the continuum model assumed,
  the disk radius is related to the square root of the disk
  normalization (e.g. measured normalizations that are a factor of
  2--4 smaller than the simulated normalization indicate a radius that
  is too small by a factor of 1.4--2.0).}
\medskip

\clearpage

\centerline{~\psfig{file=f8.ps,width=6.0in,angle=-90}~}
\figcaption[h]{\footnotesize The plot above shows XMM-Newton MOS "full
frame" and pn "timin" mode spectra of GX 339-4 (see Miller et al.\
2006, Reis et al.\ 2008, and Done and Diaz Trigo 2009).  Each spectrum
was fit with a simple absorbed disk blackbody plus power-law model,
ignoring the 4--7~keV band while fitting the continuum.  The
associated data/model ratios are shown in the bottom panel.  The MOS
spectra were extracted using inner exclusion radii of 0" (black), 18"
(red), 25" (green), and 50" (yellow), and an outer radius of 120" in
all cases.  In strong agreement with our simulations, the line profile
in the black spectrum, which includes the piled-up core, is weaker and
narrower than the line revealed in spectra extracted from annuli.  The
pn spectra were extracted using differing exclusion stripes, of 0"
(cyan), 20.5" (magenta), and 49.2" (blue).  The data/model ratio for
each pn spectrum reveals trends that match those found in our
simulations of photon pile-up.  However, the trends are not removed by
excluding the center of the PSF, which argues against a strong pile-up
contribution.  Several other effects may also contribute to the pn
residuals, including calibration uncertainties in the pn response
function which become important at this high level of sensitivity.}
\medskip

\clearpage

\centerline{~\psfig{file=f9.ps,width=4.0in,angle=-90}~}
\figcaption[h]{\footnotesize The plot above shows relativistic line
profiles detected in separate observations of the neutron star X-ray
binary Serpens X-1 using {\it Suzaku} (black) and {\it XMM-Newton}
(red).  The line is broader in the {\it Suzaku} observation, though
there is a clear red wing in the {\it XMM-Newton} line profile.
(Figure adapted from a panel in Figure 9 of Cackett et al.\ (2010).)}
\medskip

\centerline{~\psfig{file=f10.eps,width=4.0in}~}
\figcaption[h]{\footnotesize The figure above shows the relativistic
line profile observed in Serpens X-1 with {\it Suzaku}.  Photon
pile-up can be checked and mitigated by extracting annular regions
that avoid the center of the PSF.  The different spectra shown above
correspond to annuli of different inner exclusion radii: 0 pixels
(black), 30 pixels (red), 60 pixels (green), and 90 pixels (blue).  In
this particular case, the line profile is clearly consistent in all
spectra, indicating that pile-up in this spectrum is mild and does not
affect the line profile significantly.  (Figure adapted from Figure 2
in Cackett et al.\ 2010).}
\medskip

\centerline{~\psfig{file=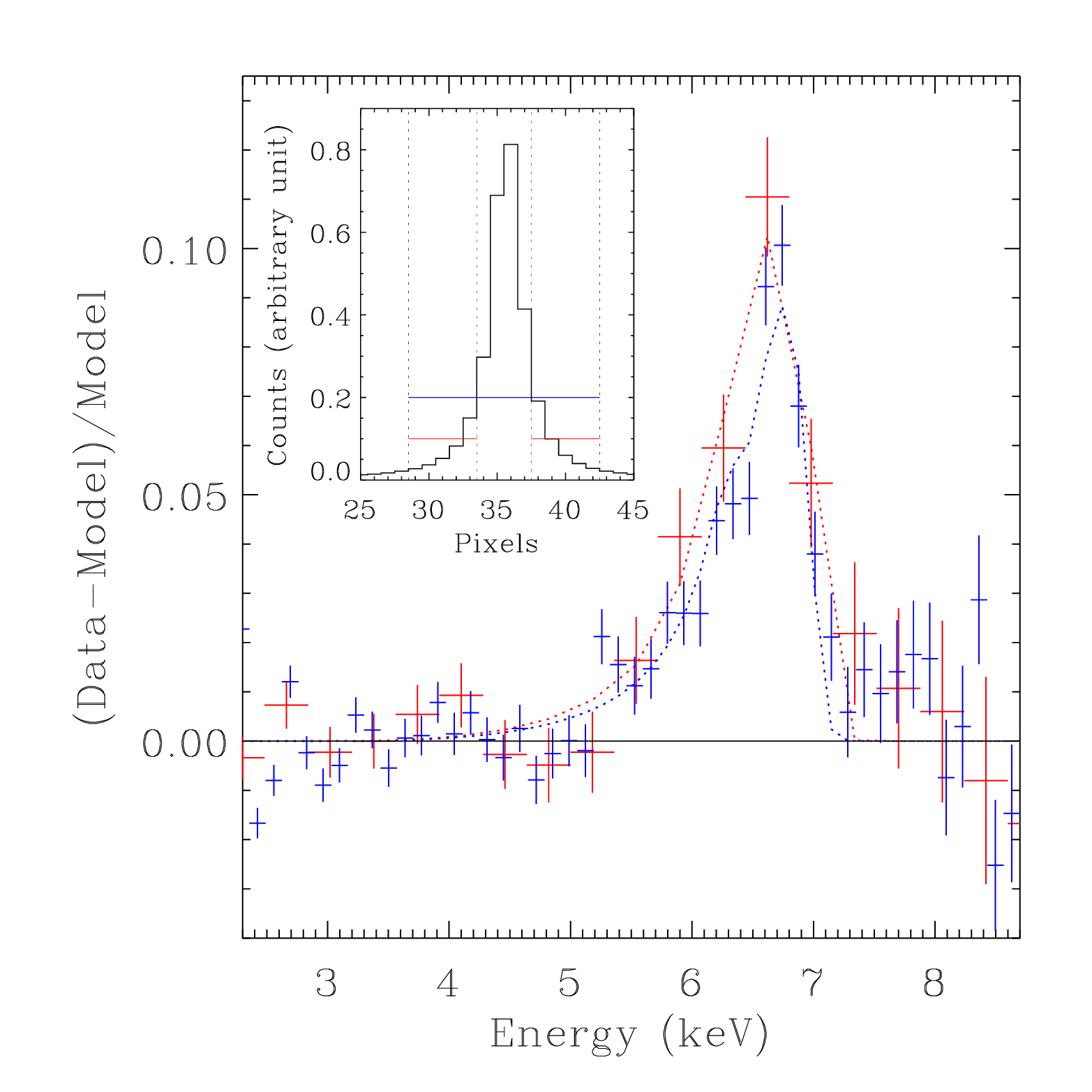,width=4.0in}~}
\figcaption[h]{\footnotesize The figure above shows the relativistic
line profile observed in Serpens X-1 with {\it XMM-Newton}.  The blue
data points and dotted model curve correspond to an extraction region
including the center of the PSF.  The red data points and dotted model
line correspond to an extraction region that excludes the brightest
central pixels.  The inset figure displays the variation in the total
counts extracted as a function of the extraction region.  The
relativistic profile of the line is evident in both spectra, and fully
consistent.  This strongly suggests that the relativistic line shape
is {\it not} due to photon pile-up, and that excluding the center of
the PSF results in a poorly-defined line profile that is statistically
more consistent with a Gaussian profile owing only to the reduced
sensitivity of the spectrum.}

\end{document}